\documentclass[showpacs,preprintnumbers,amsmath,amssymb,floatfix,nofootinbib]{revtex4-1}
\usepackage[usenames]{color}
\usepackage{epsfig}
\usepackage{amssymb}
\usepackage{multirow}
\usepackage{xcolor}
\newcommand{\be}{\begin{equation}}
\newcommand{\ee}{\end{equation}}
\newcommand{\bee}{\begin{eqnarray}}
\newcommand{\eee}{\end{eqnarray}}

\definecolor{grey}{rgb}{0.9,0.9,0.9}
\definecolor{black}{rgb}{0,0,0}

\def \irbaddress{Rudjer Bo\v{s}kovi\'{c} Institute, Bijeni\v{c}ka cesta 54, P.O. Box 180, 10002 Zagreb, Croatia}
\def \untzaddress{University of Tuzla, Faculty of Science, Univerzitetska 4, 75000 Tuzla, Bosnia and Herzegovina}
\def \mainzaddress{Institut f\"{u}r Kernphysik, Universit\"{a}t Mainz, D-55099 Mainz, Germany}
\def \GWUSAIDaddress{Data Analysis Center at the Institute for Nuclear Studies, Department
of Physics, The George Washington University, Washington, D.C. 20052}

\begin{document}

\title{Pole positions and residues from pion photoproduction using the \\ Laurent+Pietarinen expansion method}
\author{Alfred \v{S}varc}
\email{alfred.svarc@irb.hr}
\affiliation{\irbaddress}
\author{Mirza Had\v{z}imehmedovi\'{c}}
\affiliation{\untzaddress}
\author{Hedim Osmanovi\'{c}}
\affiliation{\untzaddress}
\author{Jugoslav Stahov}
\affiliation{\untzaddress}
\author{Lothar Tiator}
\affiliation{\mainzaddress}
\author{Ron L. Workman}
\affiliation{\GWUSAIDaddress}

\begin{abstract}
We have applied a new approach to determine the pole positions and residues from pion
photoproduction multipoles. The method is based on a Laurent expansion of the partial wave
T-matrices, with a Pietarinen series representing the regular part of energy-dependent and
single-energy photoproduction solutions. The method has been applied to multipole fits generated by
the MAID and GWU/SAID groups. We show that the number and properties of poles extracted from
photoproduction data correspond very well to results from $\pi$N elastic data and values cited by
Particle Data Group (PDG). The photoproduction residues provide new information for the
electromagnetic current at the pole position, which are independent of background parameterizations,
as opposed to the Breit-Wigner representation.
Finally, we present the photo-decay amplitudes from the
current MAID and SAID solutions at the pole, for all four-star nucleon resonances below $W=2$~GeV.
\end{abstract}


\date{\today}

\maketitle
\section{Introduction}
\noindent

Revisions to the Review of Particle Properties, by the Particle Data Group (PDG) \cite{PDG}, and
contributions to recent baryon spectroscopy workshops \cite{Camogli2012,Kloster2013} have
emphasized the fact that poles, and not Breit-Wigner parameters, properly determine and quantify
resonance properties linking scattering theory and QCD. However, the optimal method for extracting
pole parameters, from single-channel T-matrices, remains an open question. Experimentalists are
quite familiar with fits to data using Breit-Wigner functions (either with constant parameters and
very general backgrounds, or with energy dependent masses and widths), but are less experienced
when complex energy poles are desired. At present, poles are usually extracted from theoretical
single or multi-channel models, which are first solved with free parameters fitted to the data.
Only then is an array of standard pole extraction methods applied: analytic continuation of the
model functions into the complex energy plane \cite{Doering,EBAC,CMB,Zagreb,Bonn:2012}, speed plot
\cite{Hoehler93}, time delay \cite{Kelkar}, N/D method \cite{ChewMandelstam}, regularization
procedure \cite{Ceci2008}, etc. However, this often requires continuing an
obtained analytic solution, which implicitly contains both singular and regular (background) parts,
into the complex energy plane. Consequently, the analytic form of the full solution, and its pole
parameters, vary from model to model and the pole-background separation method requires an intimate
knowledge of the underlying model.

In Ref.~\cite{Svarc2013}  we have presented a new approach to quantifying pole parameters of single-channel processes, based on a 
Laurent expansion of partial-wave T-matrices in the vicinity of the real axis. Instead of using the conventional power-series 
description of the non-singular part of the Laurent expansion, we have represented this using a convergent series of Pietarinen 
functions.
As the analytic structure of the non-singular part is usually well understood
(including physical cuts with branch points at inelastic thresholds, and unphysical cuts in the negative energy plane), we find that 
one Pietarinen series per cut represents the analytic structure fairly reliably. The number of terms in each Pietarinen series is 
determined by the quality of the fit.  The method has been tested in two ways: on a toy model constructed from two known poles, various 
background terms, and two physical cuts, and on several sets of realistic $\pi$N elastic energy-dependent partial-wave amplitudes 
(GWU/SAID - \cite{GWU,GWU1}, and Dubna-Mainz-Taipei - \cite{DMT0,DMT}).  We have shown that the method is robust and stable, using up 
to three Pietarinen series, and is particularly  convenient in fits to single-energy solutions, which are more directly tied to 
experiment.
Apart from its ease of use, it provides a tool for the extraction and comparison of pole properties from
different analyses.  There have been several recent studies of model-dependence in single-energy
photoproduction amplitude reconstruction~\cite{ses} (both helicity and multipole),
with extensions to other reactions as well~\cite{bnga_other}. The simplicity of the our expansion method
enables us to self-consistently analyze and compare results from different approaches, as described below.

Here we have applied the new approach to determine pole positions and residues from single-pion
photoproduction multipoles.  The method has been applied to energy-dependent and single-energy
multipole fits generated by the MAID and GWU/SAID groups for eight dominant multipoles. We show
that the number and properties of poles extracted from photoproduction data correspond very well to
results from $\pi$N elastic data and values cited by Particle Data Group (PDG)~\cite{PDG}. The
photoproduction residues provide new information for the electromagnetic current at the pole
position, which is independent of background parameterizations used in the Breit-Wigner approach.

With pole positions and residues, confidently determined for MAID and SAID ED solutions, we have
further evaluated the photo-decay amplitudes at the pole positions and made a comparison with other
very recent analyses.

Below, in Section II, we give an overview of the expansion method. In Section III, this method is
applied to both energy-dependent and single-energy results from the MAID and SAID groups for eight
dominant multipoles. At this point we also discuss our extended error analysis. In Section IV, we
summarize and discuss our results for each partial wave. In section V, we present our photo-decay
amplitudes and compare them to other recent extractions.
Finally, in Section VI we summarize our results and
conclude with prospects for further work.

\section{Formalism}
For the convenience of the reader, in this Section we outline the Laurent+Pietarinen (L+P) method,
which is given a detailed description in Ref.~\cite{Svarc2013}.
\subsection{Laurent (Mittag-Leffler) expansion}

The starting point of our method is a generalization of the Laurent expansion, applied to
multipoles, using the Mittag-Leffler  theorem \cite{Svarc2013,Mittag-Leffler}, a theorem expressing
a function in terms of its $k$ first order poles and an entire function:
\begin{eqnarray}
\label{eq:Laurent}
T(W) &=& \sum _{i=1}^{k} \frac{a_{-1}^{(i)}}{W -W_i}+B^{L}(W);    \, \, \, \, a_{-1}^{(i)}, W _i, W \in  \mathbb{C}.
\end{eqnarray}

Here, $a_{-1}^{(i)}$ and $W_ i$ are residua and pole positions for the i-th pole respectively, and
$B^{L}(W )$ is a regular function in the whole complex plane. It is important to note that this
expansion is not a representation of the unknown function $T(W)$ in the full complex energy plane,
but is restricted to the part of the complex energy plane where the expansion converges, and is
defined by the area of convergence of the Laurent expansion. If we choose poles as expansion
points, the Laurent series converge on the open annulus around each pole, where the center of the
annulus is the pole position. The outer radius of the annulus extends to the position of the next
singularity (such as a nearby pole). Thus, our Laurent expansion converges on a sum of circles
located at the poles, and this part of the complex energy plane in principle includes the real
axis. Therefore, fitting the expansion (\ref{eq:Laurent}) to the experimental data on the real
axis, can in principle give the exact values of the scattering matrix poles.

The novelty of our approach is a particular choice for the non-pole contribution $B^{L}(W)$,
based on an expansion method used by Pietarinen in the context of $\pi N$ elastic scattering analysis.
Before proceeding, we briefly review this method.

\subsection{Pietarinen series}

A specific type of conformal mapping technique has been proposed and introduced by Ciulli \cite{Ciulli,Ciulli1} and Pietarinen 
\cite{Pietarinen}, and used in the Karlsruhe-Helsinki partial wave analysis \cite{Hoehler84} as an efficient expansion of
invariant amplitudes.  It was later used by a number of authors for solving various problems in scattering and field theory 
\cite{ConMap-Use}, but not applied to the pole search prior to our recent study~\cite{Svarc2013}. A more detailed discussion of the use 
of conformal mapping, and this method in particular, can be found in Refs.\cite{Svarc2013,Mittag-Leffler}.

If $F(W)$ is a general, unknown analytic function having a cut starting at $W=x_P$, then it can be represented in a power series of 
``Pietarinen functions" in the following way:

\begin{eqnarray}
\label{eq:Pietarinen}
F(W ) &=& \sum_{n=0}^{N}c_n\, X(W )^n\,, \, \, \, \, \, \, \, \, \, W  \in  \mathbb{C}\,,   \nonumber \\
X(W )&=& \frac{\alpha-\sqrt{x_P-W }}{\alpha+\sqrt{x_P-W }}\,, \, \, \, \, c_n, x_P, \alpha \in
\mathbb{R}\,,
\end{eqnarray}
with $\alpha$ and $c_n$ acting as tuning parameters and coefficients of the Pietarinen function, $X(W)$, respectively.

The essence of the approach is the fact that a set $(X(W )^n, \, n=1, \, \infty)$ forms a complete set of functions defined on the unit 
circle in the complex energy plane having a branch cut starting at $W= x_P$. The analytic form of the function is, at the beginning, 
undefined. The final form of the analytic function $F(W)$ is obtained by introducing the rapidly convergent power series with real 
coefficients, and the degree of the expansion is automatically determined in fitting the input data. In the exercise of 
Ref.\cite{Pietarinen}, as many as 50 terms were used; in the present case, covering a more narrow energy range, fewer terms are 
required.

\subsection{Application of the Pietarinen series to scattering theory}

The analytic structure of each partial wave is well known, with poles parameterizing resonant
contributions, cuts in the physical region starting at thresholds of elastic and all possible
inelastic channels, plus t-channel, u-channel and nucleon exchange contributions quantified with
corresponding negative energy cuts. However, the explicit analytic form of each cut contribution
remains to be determined. Instead of guessing the exact analytic form, we propose to use one
Pietarinen series to represent each cut, with the number of terms determined by the quality of the
fit to the input data. In principle, we have one Pietarinen series per cut, with known
branch-points $x_P, x_Q ...$, and coefficients determined by the fit to a real physical process. In
practice, we have too many cuts (especially in the negative energy range), thus we reduce their
number by dividing them in two categories: all negative energy cuts are approximated with only one,
effective negative energy cut represented by one (Pietarinen) series (we denote its branch point as
$x_P$), while each physical cut is represented by a separate series with branch-points
($x_Q,x_R...$) determined by the physics of the process. In our present analysis, we
fit all partial waves starting at the pion threshold. Therefore we fix our second branch point,
$x_Q$ to the pion threshold at $W=1077$~MeV. For the third branch point $x_R$ we will compare
results when a physical branch point is either fixed or fitted. Further branch points were found to
be unnecessary in our present data analysis.

In summary, the set of equations which define the Laurent expansion + Pietarinen series method (L+P method) is:

\begin{eqnarray}
\label{eq:Laurent-Pietarinen}
T(W ) &=& \sum _{i=1}^{k} \frac{a_{-1}^{(i)}}{W-W_i}+ B^{L}(W)
\nonumber \\
B^{L}(W)&=& \sum _{n=0}^{M}c_n\, X(W )^n  +  \sum _{n=0}^{N}d_n\, Y(W )^n +  \sum _{n=0}^{N}e_n\, Z(W )^n  + \cdots
\nonumber  \\
X(W )&=& \frac{\alpha-\sqrt{x_P-W}}{\alpha+\sqrt{x_P - W }}; \, \, \, \, \,   Y(W ) =  \frac{\beta-\sqrt{x_Q-W }}{\beta+\sqrt{x_Q-W }};  
\, \, \, \, \,   Z(W ) =  \frac{\gamma-\sqrt{x_R-W }}{\gamma+\sqrt{x_R-W }} + \cdots
\nonumber \\
&& a_{-1}^{(i)}, W _i, W   \in   \mathbb{C}
\nonumber \\
&& c_n, d_n, e_n \alpha, \beta, \gamma ... \in  \mathbb{R}  \, \,
{\rm and} \, \, x_P, x_Q, x_R  \in \mathbb{R} \, \,  {\rm  or} \, \, \mathbb{C}
\nonumber \\
&& {\rm and} \, \, \, k, M, N  ... \in  \mathbb{N}.
\end{eqnarray}

As our input data are on the real axes, the fit is performed only on this dense subset of the complex energy plane. All Pietarinen 
parameters in set of equations (\ref{eq:Laurent-Pietarinen}) are determined by the fit.

We observe that the class of input functions which may be analyzed with this method is quite extensive.
One may either fit partial-wave amplitudes obtained from theoretical models, or possibly experimental data directly. In either case, 
the T-matrix is represented by this set of equations (\ref{eq:Laurent-Pietarinen}), and minimization is usually carried out in terms of 
$\chi^2$.

\subsection{Real and complex branch points}

While the fit strategy outlined in Eqs.(3) implies the use of purely real branch points, we know
that there are, in principle, also complex branch points in the complex energy plane. This feature
can be seen simply starting from three-body unitarity conditions~\cite{Gribov}.
However, real or complex branch points  describe different physical situations. If the branch points $x_P$, $x_Q$.... are real numbers, 
this means that our background contributions are defined by stable initial and final state particles. All resonance contributions to 
the observed process are created by intermediate isobar resonances; all other initial and final state contributions are given by stable 
particles,  and are described by Pietarinen expansions with real branch point coefficients.  From experience, we know that this is not 
true: a three-body final state is always created, provided that the energy balance allows for it, and in three-body final states we 
typically do have a contribution from one stable particle (nucleon or pion), and many other combinations of two-body resonant 
sub-states, such as $\sigma$, $\rho$, $\Delta\ldots\,$. These resonant sub-states produce complex branch points.

As will be demonstrated below,
we claim that the single-channel character of the L+P method prevents us from establishing,
with certainty, which mechanism dominates. Specifically, with only single-channel information available, we have two alternatives: 
either we obtain a good fit with an extra resonance and stable initial and final state particles (real branch points), or we may obtain 
a good fit with one resonance fewer, and a complex branch point. Having only
single-channel data, we are not able to distinguish between the two. This effect was already noted,
in the context of the J\"{u}lich model, with a $\rho$N complex branch point interfering and intermixing with the $N(1710)1/2^+$ 
resonance signature, as discussed in Ref. \cite{Ceci2011}.

One advantage of the Pietarinen expansion method is its simple extension to complex branch points. We can check the above statements 
through applications of the L+P method to the photoproduction multipole $M_{1-}^{1/2}$, connected
to the $P_{11}$ partial wave from elastic pion-nucleon scattering.

\section{Application of the L+P method to pole extraction from photoproduction multipoles}

Fits to photoproduction data, particularly pion and kaon photoproduction, have been significantly
advanced with the availability of new and precise measurements of polarization observables (both
single- and double-polarization). This has revived the study of amplitude reconstruction from data
with minimal theoretical input. However, significant discrepancies do still remain in comparisons
of the major analyses. Results are generally reported either as energy-dependent (ED) fits, giving
a functional representation of the amplitudes over some extended energy range, or as single-energy
(SE) solution, which analyze data in narrow bins of energy. In the SE case, a significant variation
is possible, as a given bin does not generally contain a sufficient set of observables to uniquely
determine an amplitude. Some constraints from the underlying ED fit are usually necessary to obtain
a fit. Still, these SE fits do give a better representation of the data and can give hints of
structure possibly missing in the global ED fit. It is therefore of interest to find a method of
extracting resonance information from these, less smooth, sets of amplitudes. In the ED case, if
one has the fit function, it is in principle possible to locate poles from a mapping of the
amplitude in the complex energy plane.

In this paper we use the flexibility of the proposed L+P method (usable for both theoretical and
experimental input), to extract pole parameters (pole positions and residues) for two well-known
sets of ED and SE photoproduction amplitudes: the MAID\cite{MAID} and GWU/SAID \cite{GWU} results
for single-pion photoproduction. Electric and magnetic multipole amplitudes are analyzed in the
fits.

\subsection{The fitting procedure}

We use three Pietarinen functions (one with a branch-point in the unphysical region to represent all left-hand cuts, and two with  
branch-points in the physical region, to represent the dominant inelastic channels), combined with the minimal number of poles. In 
addition, we allow the possibility  that one of the branch points becomes a complex number, accounting for all three-body final states 
in an effective manner. We generally start with 5 Pietarinen terms per decomposition, and the anticipated number of poles (one for most 
channels, two for $E_{0+}^{1/2}$). The discrepancy criteria are defined below in terms of reduced $\chi^2_{dp}$ for SE and its analog 
- discrepancy parameter $D_{dp}$ for ED solutions. This quantity is minimized using MINUIT and the quality of the fit is also visually 
inspected by comparing the fitting function with fitted data. If the fit is unsatisfactory (discrepancy parameters are too high, or fit 
visually does not reproduce the fitted data), at first the number of Pietarinen terms is increased and, if this does not help, the 
number of poles is increased by one. The fit is repeated, and the quality of the fit is re-estimated. This procedure is continued until 
we have reached a satisfactory fit.

Pole positions, residues, and Pietarinen coefficients $\alpha$, $\beta$, $\gamma$, $c_i$, $d_i$ and $e_i$ are
our fitting parameters. However, in the strict spirit of the method, Pietarinen branch points $x_P$, $x_Q$ and $x_R$ should not be 
fitting parameters; we have declared that each known cut should be represented by its own Pietarinen series, fixed to known physical 
branch points. While this would be ideal, in practice the application is somewhat different.  We can never include all physical cuts 
from the multi-channel process. Instead, we represent them by a smaller subset. Thus, in our model, Pietarinen branch points $x_P$, 
$x_Q$ and $x_R$ are not generally constants;  we have explored the effect of allowing them to vary as fitting parameters. In the 
following, we shall demonstrate that when searched, the branch points in the physical region still naturally converge towards 
branch-points which belong to channels which dominate particular partial waves, but may not actually correspond to them exactly. The 
proximity of the fit results to exact physical branch points describes the ``goodness of the fit", namely it tells us how  well certain 
combinations of thresholds are indeed approximating a partial wave. And this, together with the choice of the degree of the Pietarinen 
polynomial, represents the model dependence of our method. We do not, of course, claim that our method is entirely model independent. 
However, the method chooses the simplest function with the given analytic properties which fits the data, and increases the complexity 
of the function only when the data require it.

\subsection{Error analysis}

In our principal paper \cite{Svarc2013} we have tested the validity of the model on a number of
well known $\pi$N amplitudes, and concluded that the method is very robust and stable. However, in
that paper we did not present an error analysis, deferring it to the forthcoming paper. We have
fulfilled this promise in Ref.~\cite{Svarc2014}, and we repeat its essence for the convenience of
the reader.

For energy-dependent solutions, we introduce the discrepancy or deviation parameter per data point
$D_{dp}$ (the substitute for $\chi^2$ per data point, $\chi^2_{dp}$,
when analyzing experimental data) in the
following way:

\begin{eqnarray} \label{def:D}
D_{dp}  &=&  \frac{1}{2 \, N_{E}} \, \, \sum_{i=1}^{N_{E}} \left[ \left( \frac{{\rm Re}
T_i^{fit}-{\rm Re} T_i^{ED}}{Err_i^{\rm Re}}  \right)^2 +    \left( \frac{{\rm Im} T_i^{fit}-{\rm
Im} T_i^{ED}}{Err_i^{\rm Im}} \right)^2 \right]\,,
\end{eqnarray}

where $N_{E}$ is the number of energies, and errors of energy-dependent solutions are introduced as:

\begin{eqnarray}  \label{def:Err}
Err_i^{\rm Re} & = &  0.05 \, \, \frac{ \sum_{k=1}^{N_{E}} |{\rm Re} T_k^{ED}|} {N_{E}} +
0.05 \, \, |{\rm Re} T_i^{ED}| \,,
\nonumber \\
Err_i^{\rm Im} & = &  0.05 \, \, \frac{ \sum_{k=1}^{N_{E}} |{\rm Im} T_k^{ED}|} {N_{E}} +
0.05 \, \, |{\rm Im} T_i^{ED}| \,. \nonumber
\end{eqnarray}
When errors of the input amplitudes are not given, and one wants to make a minimization, errors
have to be ``defined". There are two simple ways to do it: either to assign a constant error to
each data point, or introduce an energy-dependent error as a certain percentage of the given value.
However, both definitions have drawbacks. For the first recipe only high-valued points are favored,
while in the latter case low-valued points tend to be almost exactly reproduced. We find neither of
these to be satisfactory, thus we follow prescriptions chosen by the GWU and Mainz groups, and use
a sum of constant and energy dependent errors.

In the L+P method, we consider both statistical and systematic errors. Statistical errors are simply
taken over from the MINUIT program, which is used for minimization. It is shown separately in all
tables as the first term. Systematic errors are the errors of the method itself, and require a more
detailed explanation. By construction it is clear that the method has its natural limitations. Our
Laurent decomposition contains only two branch points in the physical region, and this is far from
enough in a realistic case. Functions representing the multipole amplitudes, in principle
containing more than two branch points, will in our model be approximated by a different analytic
function containing only two. This approximation will be the main source of our errors. Therefore,
we define the following procedure for quantifying systematic errors:

\begin{itemize}
               \item [i)]We completely release the first (unphysical) branch point $x_P$, as this represents
a sum of many background contributions.
               \item [ii)] We keep the first physical branch point $x_Q$ fixed at $x_Q=1077$ MeV (the $\pi$N threshold)  because we know 
that this threshold branch point should always be present.
               \item [iii)] The error analysis is done by varying the remaining physical branch point $x_R$ in two ways:
                    \begin{description}
                         \item [a] We fix the third branch point $x_R$ to the threshold of the dominant inelastic channel for the 
chosen partial wave (for instance $\eta$ threshold for S-wave) if only one inelastic channel is important, or in case of several 
equally important inelastic processes we perform several fits with the $x_R$ branch point fixed to each threshold in succession.
                         \item [b] We release the third branch point $x_R$ completely allowing MINUIT to find an effective branch point 
representing all inelastic channels. It is clear that if only one channel is dominant, the result of the fit will be very close to the 
dominant inelastic channel (see $S_{11}(_{p}E_{0+})$ ($1486^{\eta N}$  vs. $1495^{free}$) or otherwise in some effective location (see 
all other partial waves)
                    \end{description}
                    \item [iv)] We average results of the fit, and obtain a standard deviation
\end{itemize}
The list of all values for the branch point $x_R$ is given in the Appendix.
\\ \\ \noindent
The quality of our fits for the ED solutions is measured by the deviation $D_{dp}$ defined in Eqs. (\ref{def:D}) and (\ref{def:Err}).
\\ \\ \noindent
For SE solutions we use the statistical errors and obtain the standard $\chi^2_{dp}$
definition:
\begin{eqnarray}
\chi^2_{dp} &=&  \frac{1}{2 \, N_{E}} \, \, \sum_{i=1}^{N_{E}} \left[ \left( \frac{{\rm Re} T_i^{fit}-{\rm Re} T_i^{SE}}{Err_i^{\rm 
Re}}  \right)^2
  +   \left( \frac{{\rm Im} T_i^{fit}-{\rm Im} T_i^{SE}}{Err_i^{\rm Im}} \right)^2 \right]\,,
\end{eqnarray}
where $Err_i^{\rm Re}$ and $Err_i^{\rm Im}$ are standard statistical errors of the SE solutions,
real and imaginary parts respectively.

\section{Results and Discussion on Photoproduction Multipoles}

\subsection{Real branch points}

\begin{figure*}[htbp]
\includegraphics[width=0.90\textwidth]{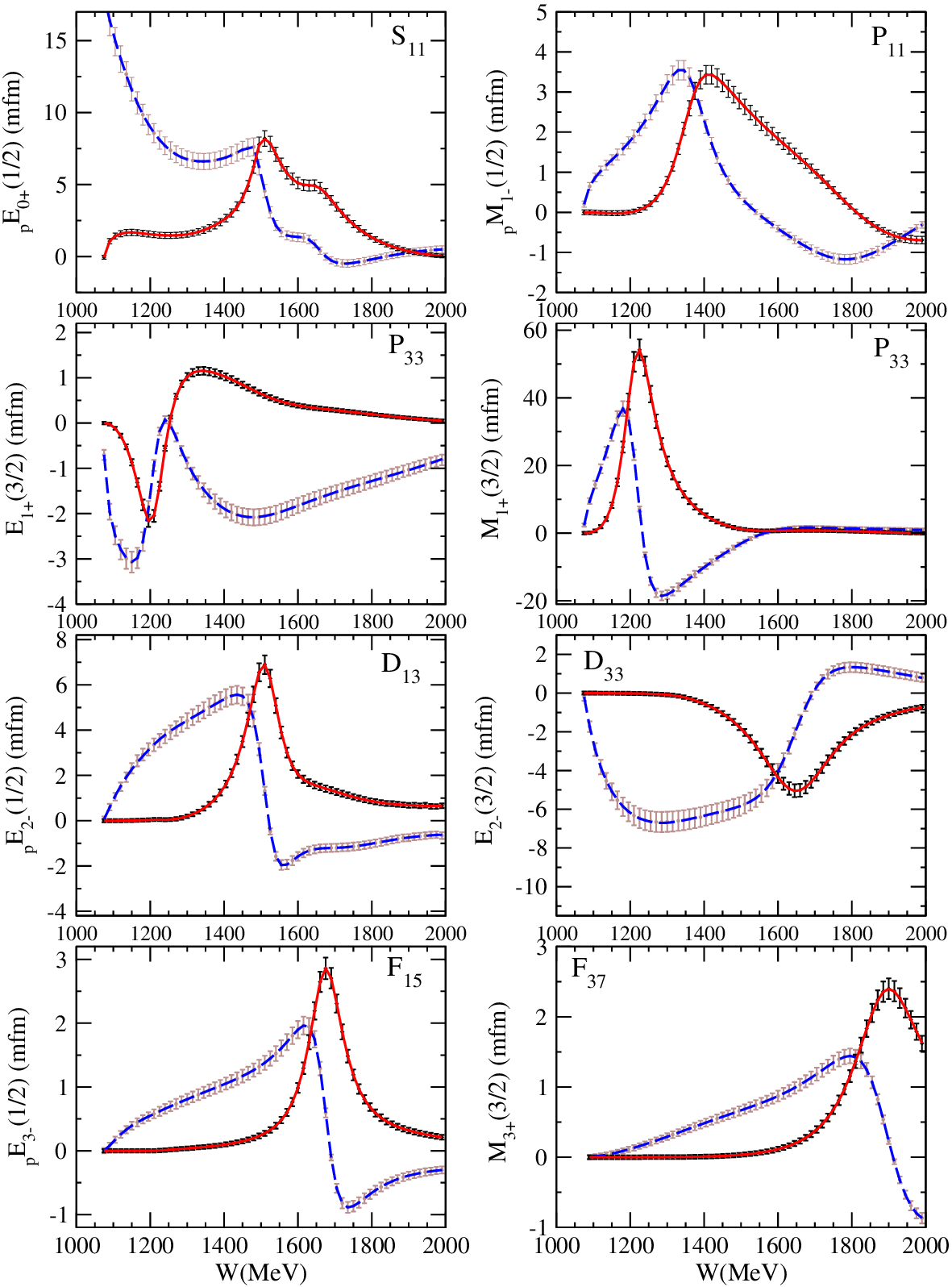}
\caption{L+P fit to GWU/SAID CM12  ED solutions. Dashed blue, and full red lines denote real and imaginary parts of multipoles 
respectively.}
\label{Fig1}
\end{figure*}

\begin{figure*}[htbp]
\includegraphics[width=0.90\textwidth]{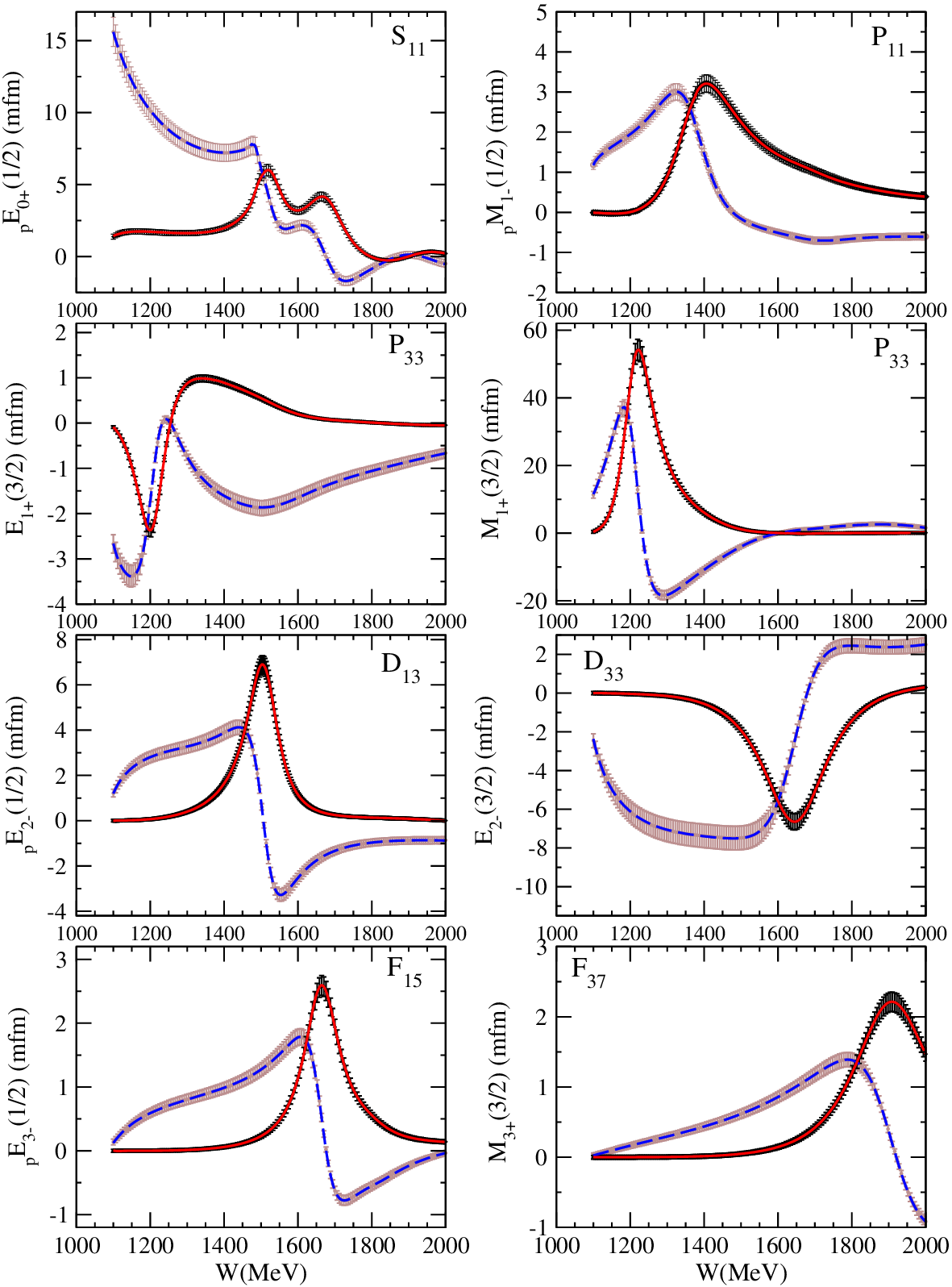}
\caption{L+P fit to MAID MAID2007 ED solutions. Dashed blue, and full red lines denote real and imaginary parts of multipoles 
respectively.}
\label{Fig2}
\end{figure*}

\begin{figure*}[htbp]
\includegraphics[width=0.9\textwidth]{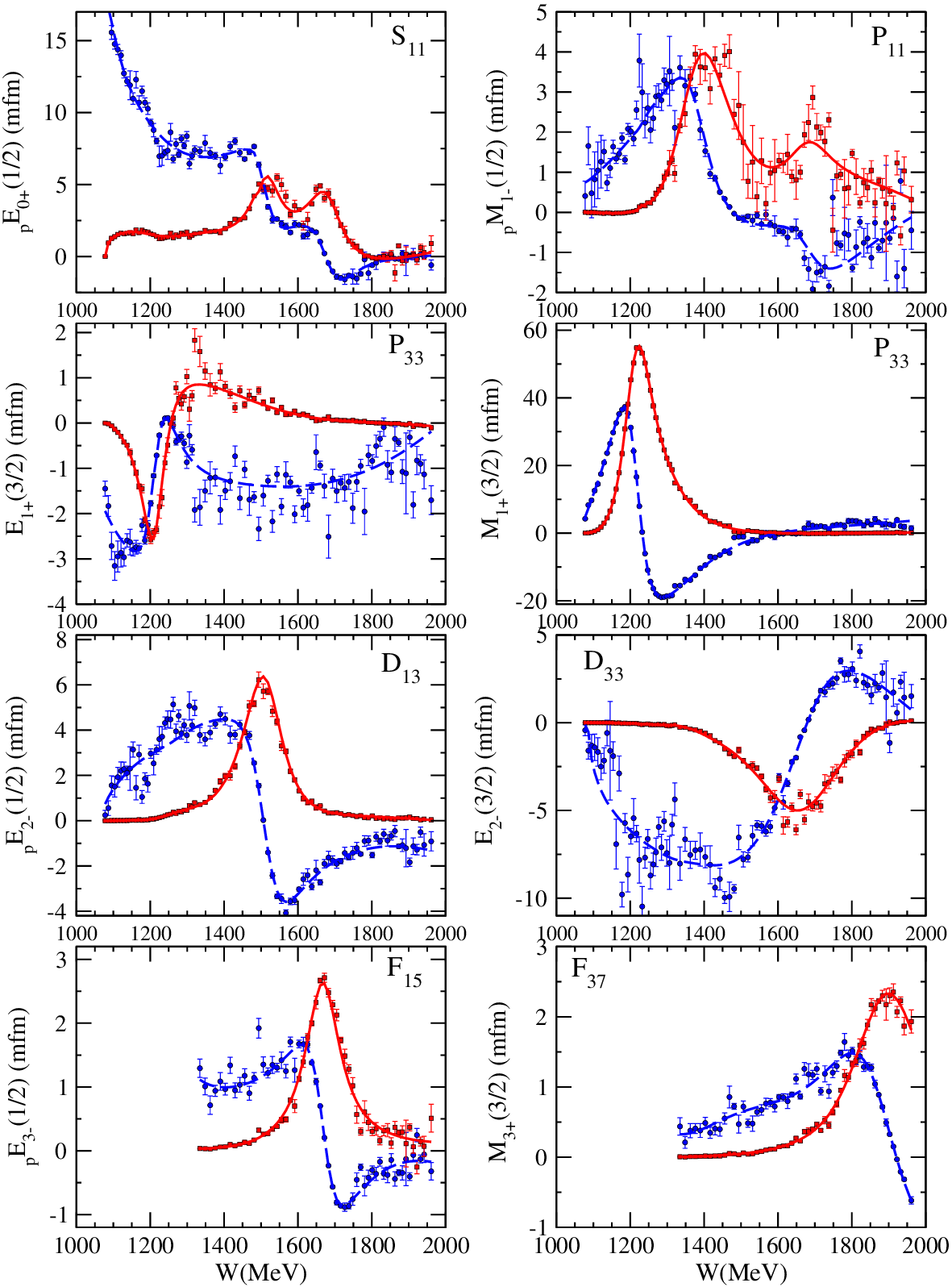}
\caption{L+P fit to MAID MAID2007 SE solutions. Dashed blue, and full red lines denote real and imaginary parts of multipoles 
respectively.}
\label{Fig3}
\end{figure*}

\begin{figure*}[htbp]
\includegraphics[width=0.9\textwidth]{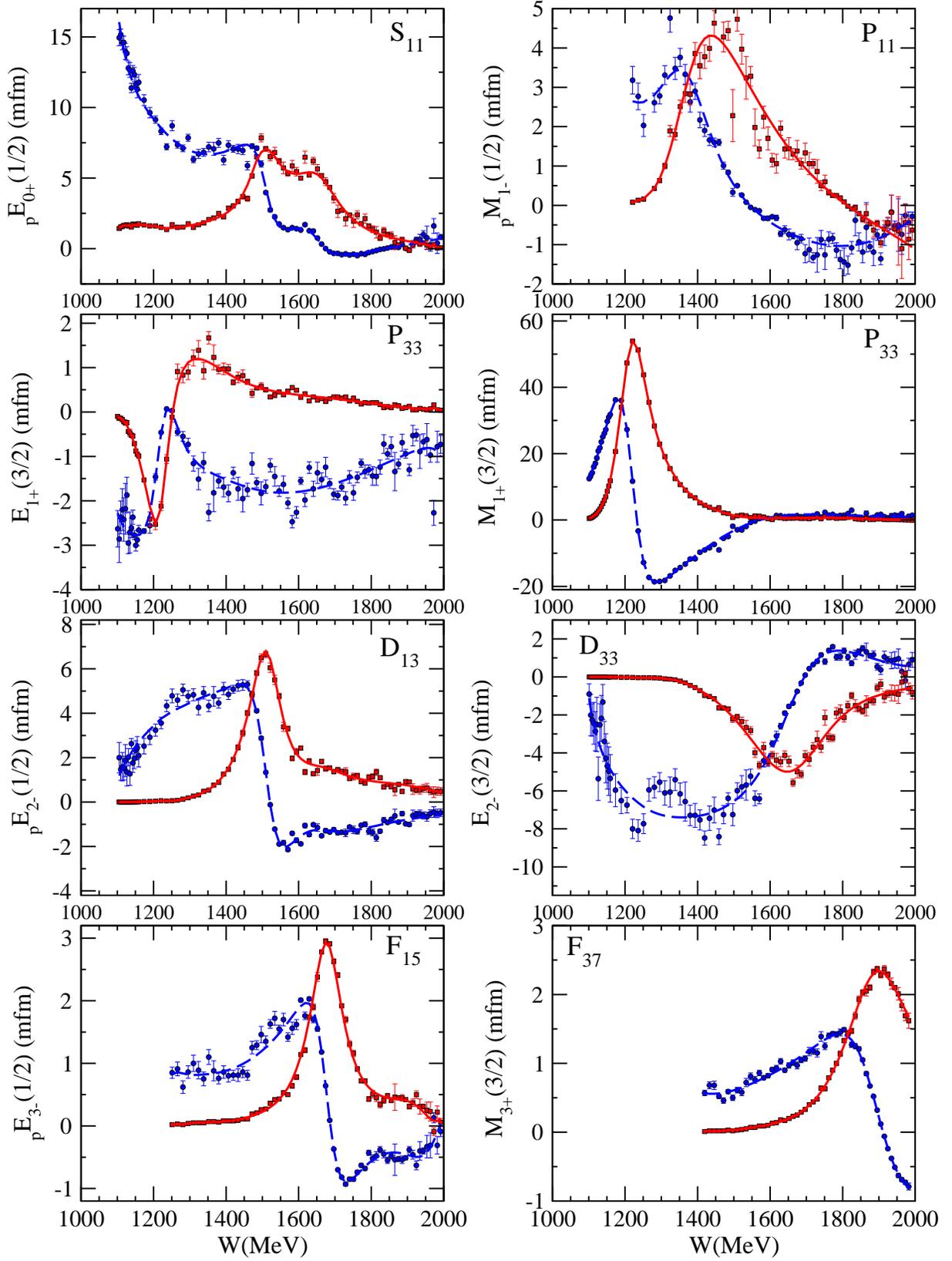}
\caption{L+P fit to GWU/SAID CM12 SE solutions. Dashed blue, and full red lines denote real and imaginary parts of multipoles 
respectively.}
\label{Fig4}
\end{figure*}

We have analyzed 24 partial wave amplitudes (electric and magnetic multipoles) from MAID and SAID
solutions up to $F$ waves. From this large number of results we have selected 8 important
multipoles for detailed discussions: $ {}_pE_{0+}^{1/2}, {}_pM_{1-}^{1/2}, E_{1+}^{3/2},
M_{1+}^{3/2}, {}_pE_{2-}^{1/2},  E_{2-}^{3/2}, {}_pE_{3-}^{1/2},  M_{3+}^{3/2}\,.$ For the rest of
the multipoles, we will present and discuss only the basic results.
\\ \\ \noindent
In Tables \ref{tab:pole1} and \ref{tab:pole2} and in Figs. \ref{Fig1} - \ref{Fig4}, we summarize
results of all our fits for real branch points. We performed the analyses for ED and SE
multipoles of the
GWU/SAID CM12  solution \cite{GWU} and the MAID MAID2007 solution \cite{MAID}.
\\ \\ \noindent
Tables \ref{tab:paramED} and \ref{tab:paramSES}, given in the Appendix, summarize the results of
the Pietarinen expansion parameters for real branch points. We show three branch points
$x_p,x_Q,x_R$ and the deviation $D_{dp}$ in case of ED solutions and the $\chi^2_{dp}$ per data
points in case of SE solutions. The latter ones should have a $\chi^2_{dp}$ close to 1. However, as
the SE points are not always Gaussian distributed data points, the $\chi^2_{dp}$ values are
generally in the range of $1-4$. For the ED solutions we have used an error definition, taking into
account relative and absolute errors of the order of $5\%$. Therefore, for a good fit, the
deviation $D_{dp}$ is much smaller, in the range of $10^{-4} - 10^{-2}$. As already discussed in
Sect. B, we have estimated the stability of our fits and the variation of the resonance parameters
by applying 3 or 4 different assumptions for the effective $3rd$ branch point.
\\ \\ \noindent
In Tables \ref{tab:pole1} and \ref{tab:pole2}, we present our results on the pole parameters of the
nucleon resonances $N^*$ and $\Delta$ that we found in our analysis. These are the pole positions
$W_p$ with $M_p=\mbox{Re}\, W_p$ and $\Gamma_p=-2\,\mbox{Im}\, W_p$ as well as $( \gamma,\pi )$
residues in terms of magnitude and phase, $R_{\gamma,\pi}=|R_{\gamma,\pi}|\,e^{i\theta}$.
Note, that the photoproduction residues, listed here, are not the residues of a $( \gamma,\pi )$
$T$-matrix, but residues of the electromagnetic multipoles $E_{\ell\pm}$ and $M_{\ell\pm}$, which
carry a dimension, e.g. mfm. Therefore, we use mfm~GeV as a convenient dimension of
$R_{\gamma,\pi}$.
\\ \\ \noindent
In the following, we discuss our results in detail for each partial wave:
\\ \\ \noindent
The $S_{11}$ partial wave is the only case with two 4-star resonances. Both resonances are well
determined from the ${}_pE_{0+}^{1/2}$ multipoles of the MAID and SAID analyses. Only for the
second state, $N(1650)1/2^-$, we find a discrepancy in the strength, it appears two times as strong
in the MAID analysis compared to SAID. A third resonance state, $N(1895)1/2^-$, is found, but only
in the MAID ED solution. It shows up with a normalized strength of $2.5$, a rather large value for
a resonance only listed with $2$-star. It will be an important candidate to watch for in
forthcoming new SE analyses from complete experiment studies.
\\ \\ \noindent
The $P_{11}$ partial wave shows a consistency only for the existence of the Roper state
$N(1440)1/2^+$. The second resonance state $N(1710)1/2^+$ (state with 3 PDG stars) is more
problematic. It varies considerably in our analyses. The width differs by a factor 5 while the
residue strength differs even more. In the MAID SE solution, see Fig.~3, a clear enhancement is
seen in the imaginary part of the ${}_pM_{1-}^{1/2}$ multipole, near 1700 MeV. In the same region,
however, the SAID SE solution in Fig.~4 appears rather smooth. This is another important case to be
better determined with future double-polarization experiments. At this point, it is worth noting,
that this state may also be compensated by different background parameterizations which both can
similarly well describe the fitted data. One possible explanation for this problem can be given
within the framework of  L+P expansion method.  Real branch points in the L+P method describe the
situation when only two-body $\rightarrow$ two-body processes contribute,  while the complex branch
point is a mathematical implementation of the situation when the three-body final state, containing
a two-body resonant sub-channel, is also important. When only real branch points are considered,
this second $N(1710)1/2^+$ state is needed to explain the data. However, when a complex $\rho N$
branch point is used (indicating a resonance in the $\pi\pi$ sub-channel of a three-body final
state), the second resonance in the $\pi N$ channel is no longer needed. This will be further
explored in the following subsection.

\begin{table*}[htbp]
{\footnotesize \caption{{\footnotesize \label{tab:pole1} Pole positions in MeV and residues of four
dominant isospin 1/2 multipoles as moduli in mfm$\,\cdot\,$GeV and phases in degrees for real
branch points. The results from L+P expansion are given for GWU/SAID and MAID energy-dependent (ED)
and single-energy (SE) solutions. Resonances marked with a star indicate resonances which can be
alternatively explained by the $\rho$N complex branch point. Empty lines indicate that a resonance
pole could not be found with a significant statistical weight. }} }{\footnotesize
\par}
\begin{tabular}{cl|lllll}
\hline
Multipole  & Source   & \textcolor{black}{Resonance}  & \textcolor{black}{$\;\mbox{Re}\, W_{p}\;$ }  & 
\textcolor{black}{$\;-2\mbox{Im}\, W_{p}\;$ }  & \textcolor{black}{$|\mbox{residue}|$ }  & \textcolor{black}{$\theta$ }\tabularnewline 
\hline \hline
\multirow{12}{*}{ $S_{11}(_{p}E_{0+})$ } & SAID  ED  & $N(1535)\:1/2^{-}$ & $1501\pm4\pm2$ & $95\pm9\pm2$ & $0.245\pm0.030\pm0.008$ & 
$-(25\pm7\pm3)^{\circ}$ \tabularnewline
 & \multicolumn{1}{c|}{MAID ED  } &  & $1516\pm 1\pm 2$ & $94\pm3\pm 2$ & $0.234\pm0.009\pm0.004$ & $-(2\pm3\pm7)^{\circ}$  
\tabularnewline
 & MAID SE   &  & $1511\pm1\pm6$ & $93\pm2\pm7$ & $0.210\pm0.002\pm0.021$ & $-(5\pm1\pm7)^{\circ}$  \tabularnewline
 & SAID SE &  & $1501\pm1\pm2$ & $112\pm2\pm7$ & $0.312\pm0.003\pm0.022$ & $-(18\pm1\pm3)^{\circ}$ \tabularnewline \cline{2-7}
 & SAID  ED  & $N(1650)\:1/2^{-}$ & $1655\pm8\pm3$ & $127\pm10\pm7$ & $0.119\pm0.019\pm0.013$ & $-(18\pm14\pm9)^{\circ}$  
\tabularnewline
 & MAID ED   &  & $1678\pm2\pm2$ & $135\pm3\pm2$ & $0.289\pm0.010\pm0.009$ & $+(12\pm3\pm4)^{\circ}$  \tabularnewline
 & MAID SE   &  & $1681\pm1\pm3$ & $113\pm1\pm6$ & $0.231\pm0.001\pm0.024$ & $-(21\pm1\pm6)^{\circ}$  \tabularnewline
 & SAID SE &  & $1650\pm1\pm1$ & $117\pm2\pm14$ & $0.153\pm0.002\pm0.026$ & $-(8\pm5\pm5)^{\circ}$  \tabularnewline \cline{2-7}
 & SAID  ED  &  $N(1895)\:1/2^{-}$ & - & - & - & -\tabularnewline
 & MAID ED   & & $1913\pm4\pm8$ & $258\pm10\pm37$ & $0.327\pm 0.015 \pm 0.2$ & $-(68\pm4\pm10)^{\circ}$  \tabularnewline
 & MAID SE   &  & - & - & - & -\tabularnewline
 & SAID SE &  & - & - & - & -\tabularnewline
\hline\hline
\multirow{8}{*}{$P_{11}(_{p}M_{1-})$ } & SAID  ED  & $N(1440)\:1/2^{+}$ & $1360\pm4\pm1$ & $183\pm10\pm9$ & $0.290\pm0.015\pm0.039$ & 
$-(61\pm4\pm1)^{\circ}$ \tabularnewline
 & MAID ED   &  & $1367\pm1\pm1$ & $190\pm3\pm2$ & $0.306\pm0.011\pm0.004$ & $-(44\pm4\pm1)^{\circ}$ \tabularnewline
 & MAID SE   &  & $1379\pm2\pm4$ & $183\pm3\pm5$ & $0.394\pm0.003\pm0.005$ & $-(36\pm1\pm5)^{\circ}$  \tabularnewline
 & SAID SE &  & $1367\pm2\pm8$ & $235\pm3\pm8$ & $0.547\pm0.006\pm0.052$ & $-(75\pm1\pm6)^{\circ}$  \tabularnewline \cline{2-7}
 & SAID  ED  & $N(1710)^*\:1/2^{+}$ & $1789\pm9\pm4$ & $550\pm25\pm3$ & $0.609\pm0.031\pm0.014$ & $+(98\pm3\pm4)^{\circ}$  
\tabularnewline
 & MAID ED   &  & $1694\pm22\pm12$ & $269\pm44\pm35$ & $0.029\pm0.005\pm0.008$ & $+(65\pm5\pm9)^{\circ}$   \tabularnewline
 & MAID SE   &  & $1678\pm5\pm3$ & $99\pm14\pm23$  & $0.062\pm0.006\pm0.012$ & $-(16\pm4\pm2)^{\circ}$   \tabularnewline
 & SAID SE &  & - & - & - & -\tabularnewline
\hline \hline
\multirow{8}{*}{$D_{13}(_{p}E_{2-})$} & SAID  ED  & \textcolor{black}{$N(1520)\;3/2^{-}$} & $1514\pm1\pm0$ & $109\pm4\pm1$ & 
$0.373\pm0.017\pm0.010$ & $+(16\pm2\pm1)^{\circ}$  \tabularnewline
 & MAID ED   &  & $1509\pm1\pm0$ & $106\pm1\pm1$ & $0.375\pm0.003\pm0.001$ & $+(11\pm1\pm1)^{\circ}$  \tabularnewline
 & MAID SE   &  & $1514\pm1\pm4$ & $120\pm1\pm6$ & $0.385\pm0.005\pm0.024$ & $+(12\pm1\pm2)^{\circ}$ \tabularnewline
 & SAID SE &  & $1514\pm1\pm1$ & $111\pm1\pm0.5$ & $0.382\pm0.004\pm0.003$ & $+(14\pm1\pm3)^{\circ}$ \tabularnewline \cline{2-7}
 & SAID  ED  & \textcolor{black}{$N(1700)^*\;3/2^{-}$} & $1638\pm13\pm13$ & $362\pm24\pm17$ & $0.382\pm0.032\pm0.059$ & 
$+(4\pm5\pm11)^{\circ}$\tabularnewline
 & MAID ED   &  & - & - & - & -\tabularnewline
 & MAID SE   &  & - & - & - & -\tabularnewline
 & SAID SE   &  & $1654\pm5\pm15$ & $257\pm10\pm47$ & $0.187\pm0.007\pm0.080$ & $-(1\pm3\pm7)^{o}$  \tabularnewline
\hline \hline
\multirow{8}{*}{$F_{15}(_{p}E_{3-})$} & SAID  ED  & \textcolor{black}{$N(1680)\;5/2^{+}$} & $1674\pm2\pm0.5$ & $113\pm4\pm0$ & 
$0.157\pm0.008\pm0$ & $-(5\pm3\pm0)^{\circ}$\tabularnewline
 & MAID ED   &  & $1663\pm1\pm0$ & $118\pm2\pm1$ & $0.150\pm0.003\pm0.001$ & $-(3\pm1\pm1)^{\circ}$ \tabularnewline
 & MAID SE   &  & $1669\pm1\pm1$ & $113\pm1\pm1$ & $0.145\pm0.005\pm0.002$ & $+(2\pm1\pm1)^{\circ}$ \tabularnewline
 & SAID SE &  & $1677\pm1\pm1$ & $115\pm1\pm3$ & $0.174\pm0.002\pm0.008$ & $+(1\pm1\pm2)^{\circ}$ \tabularnewline \cline{2-7}
 & SAID  ED  & \textcolor{black}{$N(2000)^*\;5/2^{+}$} & - & - & - & -\tabularnewline
 & MAID ED   &  & $1801\pm14\pm4$ & $141\pm28\pm13$ & $0.007\pm0.002\pm0.003$ & $+(32\pm14\pm9)^{\circ}$   \tabularnewline
 & MAID SE   &  & - & - & - & -\tabularnewline
 & SAID SE   &  & $1923\pm4\pm68$ & $172\pm30\pm22$ & $0.081\pm0.004\pm0.047$ & $+(62\pm3\pm87)^{\circ}$ \tabularnewline \hline \hline
\end{tabular}
\end{table*}


\begin{table*}[htbp]
{\footnotesize \caption{{\footnotesize \label{tab:pole2} Pole positions in MeV and residues of four
dominant isospin 3/2 multipoles as moduli in mfm$\,\cdot\,$GeV and phases in degrees for real
branch points. The results from L+P expansion are given for GWU/SAID and MAID energy-dependent (ED)
and single-energy (SE) solutions. Empty lines indicate that a resonance pole could not be found
with a significant statistical weight. }} }
\begin{tabular}{cl|lllll} \hline
Multipole  & Source  & \textcolor{black}{Resonance}  & \textcolor{black}{$\;\mbox{Re}\, W_{p}\;$}  & \textcolor{black}{$\;-2\mbox{Im}\, 
W_{p}\;$}  & \textcolor{black}{$|\mbox{residue}|$}  & \textcolor{black}{$\theta$} \tabularnewline \hline \hline
\multirow{8}{*}{$P_{33}(E_{1+})$ } & SAID  ED  & \textcolor{black}{$\Delta(1232)\;3/2^{+}$ } & $1211\pm0.5\pm1$ & $101\pm1\pm0$ & 
$0.183\pm0.005\pm0.001$ & $-(154\pm1\pm1)^{\circ}$ \tabularnewline
 & MAID ED  &  & $1211\pm0.5\pm0.5$ & $99\pm0.5\pm0.5$ & $0.184\pm0.002\pm0.003$ & $-(155\pm1\pm1)^{\circ}$  \tabularnewline
 & MAID SE  &  & $1215\pm0\pm4$ & $87\pm0\pm1$ & $0.154\pm0.001\pm0.010$ & $-(155\pm0\pm8)^{\circ}$ \tabularnewline
 & SAID SE &  & $1220\pm1\pm1$ & $85\pm1\pm2$ & $0.146\pm0.002\pm0.002$ & $-(143\pm1\pm1)^{\circ}$ \tabularnewline \cline{2-7}
 & SAID  ED  & \textcolor{black}{$\Delta(1600)\;3/2^{+}$ } & $1470\pm16\pm15$ & $396\pm34\pm17$ & $0.127\pm0.099\pm0.014$ & 
$+(109\pm5\pm15)^{\circ}$  \tabularnewline
 & MAID ED  &  & $1550\pm7\pm4$ & $347\pm12\pm29$ & $0.087\pm0.005\pm0.019$ & $+(127\pm5\pm4)^{\circ}$  \tabularnewline
 & MAID SE  &  & - & - & - & -  \tabularnewline
 & SAID SE  &  & - & - & - & -  \tabularnewline
\hline \hline
\multirow{8}{*}{$P_{33}(M_{1+})$ } & SAID  ED  & \textcolor{black}{$\Delta(1232)\;3/2^{+}$ } & $1211\pm0.5\pm0.5$ & $101\pm1\pm1$ & 
$2.974\pm0.013\pm0.028$ & $-(26\pm1\pm1)^{\circ}$  \tabularnewline
 & MAID ED  &  & $1209\pm0.5\pm0.5$ & $99\pm0.5\pm0.5$ & $2.963\pm0.021\pm0.040$ & $-(31\pm1\pm1)^{\circ}$  \tabularnewline
 & MAID SE  &  & $1210\pm0\pm1$ & $100\pm0\pm1$ & $3.010\pm0.003\pm0.020$ & $-(30\pm0\pm1)^{\circ}$ \tabularnewline
 & SAID SE &  & $1211\pm0\pm0.5$ & $101\pm0\pm1$ & $3.008\pm0.002\pm0.033$ & $-(27\pm0\pm1)^{\circ}$ \tabularnewline \cline{2-7}
 & SAID  ED  & \textcolor{black}{$\Delta(1600)\;3/2^{+}$}  & $1522\pm12\pm7$ & $409\pm24\pm11$ & $1.195\pm0.100\pm0.104$ & 
$-(132\pm2\pm6)^{\circ}$  \tabularnewline
 & MAID ED  &  & $1498\pm10\pm22$ & $326\pm20\pm20$ & $0.499\pm0.005\pm109$ & $-(149\pm1\pm20)^{\circ}$   \tabularnewline
 & MAID SE  &  & - & - & - & -  \tabularnewline
 & SAID SE  &  & $1512\pm3\pm14$ & $408\pm5\pm39$ & $1.173\pm0.016\pm0.205$ & $-(144\pm1\pm9)^{\circ}$  \tabularnewline \hline \hline
\multirow{4}{*}{$D_{33}(E_{2-})$} & SAID  ED  & \textcolor{black}{$\Delta(1700)\;3/2^{-}$ } & $1650\pm4\pm0$ & $255\pm8\pm3$ & 
$0.672\pm0.026\pm0.022$ & $-(177\pm2\pm1)^{\circ}$  \tabularnewline
 & MAID ED  &  & $1649\pm1\pm1$ & $223\pm2\pm2$ & $0.874\pm0.004\pm0.011$ & $-(175\pm1\pm2)^{\circ}$  \tabularnewline
 & MAID SE  &  & $1671\pm0\pm10$ & $376\pm1\pm6$ & $1.792\pm0.001\pm0.169$ & $-(160\pm1\pm8)^{\circ}$  \tabularnewline
 & SAID SE  &  & $1662\pm1\pm3$ & $324\pm2\pm6$ & $1.075\pm0.010\pm0.044$ & $-(161\pm1\pm3)^{\circ}$ \tabularnewline
\hline \hline
\multirow{4}{*}{$F_{37}(M_{3+})$} & SAID  ED  & \textcolor{black}{$\Delta(1950)\;7/2^{+}$ } & $1884\pm3\pm1$ & $231\pm8\pm1$ & 
$0.278\pm0.016\pm0.003$ & $-(16\pm2\pm1)^{\circ}$  \tabularnewline
 & MAID ED  &  & $1898\pm1\pm1$ & $271\pm3\pm1$ & $0.339\pm0.009\pm0.003$ & $-(11\pm1\pm1)^{\circ}$ \tabularnewline
 & MAID SE&  & $1880\pm1\pm7$ & $240\pm1\pm9$ & $0.283\pm0.003\pm0.033$ & $-(24\pm1\pm6)^{\circ}$  \tabularnewline
 & SAID SE  &  & $1882\pm1\pm1$ & $236\pm2\pm2$ & $0.283\pm0.003\pm0.004$ & $-(17\pm1\pm1)^{\circ}$ \tabularnewline
\hline \hline
\end{tabular}
\end{table*}

\newpage


\clearpage

For the $P_{33}$ partial wave, we have two multipoles $E_{1+}^{3/2}$ and $M_{1+}^{3/2}$. The pole of
the $\Delta(1232)3/2^+$ shows up quite consistently. Only for the widths and the residues in the SE
analysis we obtain $10\%$ lower values. We also found the second state in this partial wave, the
$\Delta(1600)3/2^+$ in both ED solutions with some larger deviations in the $M_{1+}$ analysis. It
is remarkable, that this resonance is found in the MAID ED solution, where it is not explicitly
included in terms of a Breit-Wigner resonance, due to its status of only 3-star. However, due to
the unitarization procedure in MAID, it is implicitly contained through the $\pi N$ unitarization
phase. For the $\Delta(1232)3/2^+$ resonance, pole positions and residues were already published in
the late 90s. The numerical values which we find here with the L+P method agree very well with the
pole positions and residues from $E_{1+}^{3/2}$ and $M_{1+}^{3/2}$ amplitudes in
Refs.~\cite{Hanstein:1996,Workman:1999}.
\\ \\ \noindent
The $D_{13}$ partial wave can be analyzed in two multipoles, where the largest one, the
${}_pE_{2-}^{1/2}$ is presented here. The first state, $N(1520)3/2^-$ is very consistent in both ED
and SE analyses, the second $N(1700)3/2^-$ is only found in the SAID solutions.
\\ \\ \noindent
The $D_{33}$ partial wave is also very important in pion photoproduction, but the photo-decay
amplitudes in the Breit-Wigner parameterizations differ substantially in the PDG listings. The
figures of the $E_{2-}^{3/2}$ multipoles appear very similar for MAID and SAID solutions, while the
$\Delta(1700)3/2^-$ pole parameters found in our L+P expansion give a rather consistent picture.
However, systematic differences between the ED and SE solutions appear much larger than the
differences between MAID and SAID solutions. The newly analyzed double-polarization data of pion
photoproduction will certainly tighten constraints for this state. It is worth mentioning that some
structure is observed in the SE solutions of MAID and SAID around a c.m. energy of 1300 MeV, a
region, where certainly no resonance is expected in this partial wave. While it looks up as a peak
in the SAID solution, in MAID it appears more as a largely scattered region. Our L+P formalism
cannot find any physical explanation for this structure.
\\ \\ \noindent
The $F_{15}$ partial wave is very similar to the previous one. In the electric ${}_pE_{3-}^{1/2}$
multipole, a very pronounced resonance structure shows up for the $N(1680)5/2^+$ state and all
resonance parameters are consistently found. The second state, $N(2000)5/2^+$, another candidate
for a complex $\rho N$ branch point, shows up very inconclusively in our L+P analysis. We can find
it in the ED solution of MAID and in the SE solution of SAID. In the other two solutions it is not
seen. The parameters clearly cannot really be considered as for the same resonance state; even
the mass differs by more than 100~MeV, and the residue strength by more than an order of
magnitude. The 2012 PDG
tables list two states with a 2-star rating, $N(1860)5/2^+$ and $N(2000)5/2^+$. Further efforts are
necessary to clarify these resonances in pion photoproduction.
\\ \\ \noindent
The $F_{37}$ partial wave, finally, appears rather clean both in the figures and in the resonance
parameters of the $\Delta(1950)7/2^+$ state.
\\ \\ \noindent

\subsection{Complex branch points}

In an alternative approach, we have replaced the third real branch point $x_R$ by a complex $\rho N$
branch point, \mbox{$x_R=1708-70 \, i$}. For the $P_{11}, D_{13}$ and $F_{15}$ partial waves, where
we already discussed problems with the second resonance states, we have found solutions that can
equally well describe the partial wave data.

\begin{table*}[htbp]
 {\footnotesize
 \caption{{\footnotesize \label{tab:complex residues} Pole positions in MeV and residues
as moduli in mfm$\,\cdot\,$GeV and phases in degrees for a $\rho N$ complex branch point. The
results from the L+P expansion are given for GWU/SAID and MAID energy-dependent (ED) and
single-energy (SE) solutions.}}
 }{\footnotesize \par}
\begin{tabular}{cc|cccccccc}
\hline
Multipole  & Source  & \textcolor{black}{Resonance}  & \textcolor{black}{$\;\mbox{Re}\, W_{p}\;$ }  & 
\textcolor{black}{$\;-2\mbox{Im}\, W_{p}\;$ }  & \textcolor{black}{$|\mbox{residue}|$ }  & \textcolor{black}{$\theta$ }  &  &  &  $\, 
\, D_{dp}/\chi^{2}_{dp}$ \tabularnewline

\hline \hline $P_{11}(_{p}M_{1-})$ & SAID ED  & $N(1440)\:1/2^{+}$  & $1361$  & $192$  & $0.326$  &
$-60^{\circ}$ & &  &  $0.0051$\tabularnewline
 & MAID ED  &  & $1367$  & $188$  & $0.297$  & $-44^{\circ}$   &  &  &  $0.0043$\tabularnewline
 & MAID SE  &  & $1381$  & $178$  & $0.369$  & $-31{}^{\circ}$ & &  & $3.13$\tabularnewline
\hline \hline
$D_{13}(_{p}E_{2-})$
 & SAID ED  & \textcolor{black}{$N(1520)\;3/2^{-}$}  & $1514$  & $109$  & $0.371$  & $+16{}^{\circ}$ &  &    & $0.0078$\tabularnewline
 & SAID SE  &  & $1511$  & $108$  & $0.354$  & $+10{}^{\circ}$ &  &   & $2.63$ \tabularnewline

\hline \hline
$F_{15}(_{p}E_{3-})$
 & MAID ED  & \textcolor{black}{$N(1680)\;5/2^{+}$}  & $1662$  & $122$  & $0.125$  & $-6{}^{\circ}$ &  &   & $0.0005$\tabularnewline
 & SAID SE  &  & $1672$  & $117$  & $0.177$  & $-3{}^{\circ}$ &  &  & $3.51$ \tabularnewline

\hline \hline \\ [2.5ex]
\end{tabular}
\end{table*}

These results are shown in Table \ref{tab:complex residues}. As the deviations of the fits
$\chi^2_{dp}$ (D$_{dp}$) are almost identical in all cases, the fits to the data overlap on
Figs.~\ref{Fig1}-\ref{Fig4}, so we do not show extra figures. The only way to distinguish between
the two options (real vs. complex branch point) is a comparison with existing data in the
three-body channel, but they have not yet been included in an analysis of the type discussed in the
present study. This demonstrates quite well that either much more precise data on polarization
experiments are required and/or data in different channels, as $\gamma p\rightarrow \pi\pi N$ or
$\gamma p\rightarrow K\Lambda$ are badly needed in order to determine whether the resonance is
formed in the two-body subsystem of a three body final state (complex Pietarinen branch point), or
it is a genuine intermediate state resonance (real Pietarinen branch point).

\begin{figure}[!h]
\includegraphics[width=0.40\textwidth]{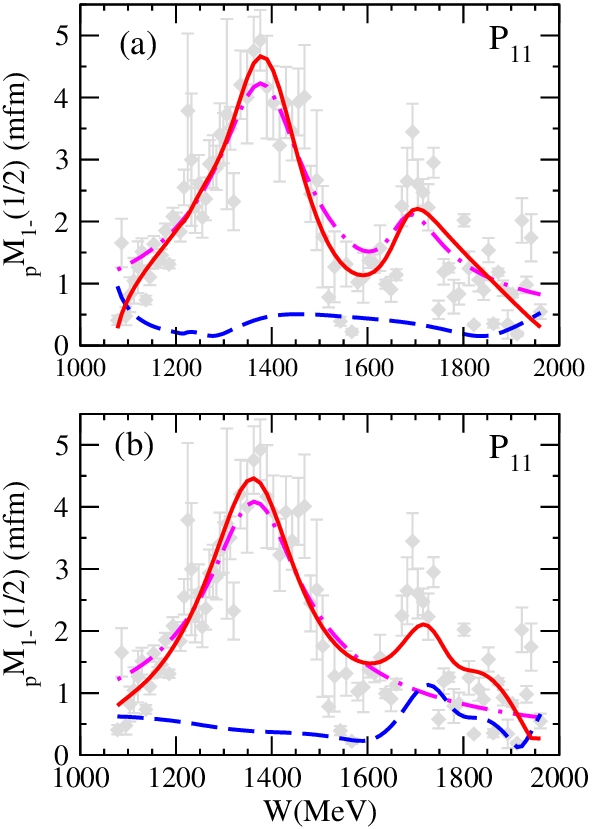}
\caption{Absolute values of the total amplitude, the resonance and background terms for the $P_{11}
(_{p}M_{1-})$ MAID SE solution are denoted by solid (red), dash-dotted (magenta) and dashed (blue)
lines respectively. Figure (a) shows the result for two resonances and a real branch point, and
Figure (b) shows the results for one resonance and a complex $\rho N$ branch point.} \label{Fig5}
\end{figure}

Just as an illustration that resonance-background separation for two-body and three-body
intermediate states is very different, in Fig.~\ref{Fig5} we have compared absolute values of
background and resonance contributions for the very important $P_{11} (_{p}M_{1-})$ multipole.

We see that the background term contains much more structure in the complex branch point case but,
lacking more data, we have to conclude that the alternative explanations are equally valid without
including another channel explicitly.

\section{Results and Discussion on Photo-decay Amplitudes}

In addition to the eight selected multipoles, which are shown in the figures and discussed above in
details with respect to single energy solutions, we have also analyzed all other multipoles from
the MAID and SAID ED solutions up to $L=3$. In order to evaluate the photo-decay amplitudes of all
13 four-star resonances below $W=2$~GeV, we also require the pole positions and residues from all
other multipoles. In Table~\ref{tab:pole}, we list all of these pole parameters needed for further
calculations together with discrepancy parameter $D_{dp}$ which is indicating the high quality of
the fit.

\begin{table*}[!bh]
{\footnotesize \caption{{\footnotesize \label{tab:pole} Pole positions in MeV and residues
of multipoles as moduli in mfm$\,\cdot\,$GeV and phases in degrees of all multipoles needed to obtain photo-decay amplitudes.
 }}
}{\footnotesize \par}
\begin{tabular}{cc|cccccc}
\hline
Multipole  & Source   & \textcolor{black}{Resonance}  & \textcolor{black}{$\;\mbox{Re}\, W_{p}\;$ }  & 
\textcolor{black}{$\;-2\mbox{Im}\, W_{p}\;$ }  & \textcolor{black}{$|\mbox{residue}|$ }  & \textcolor{black}{$\theta$ } & { 
10$^2$}{$D_{dp}$ } \tabularnewline
\hline
\hline
 $S_{31}(_{}E_{0+})$  & SAID ED & $N(1620)\:1/2^{+}$ & $1596\pm3\pm1$ & $124\pm6\pm1$ & $0.332\pm0.019\pm0.002$ & 
$(138\pm3\pm5)^{\circ}$ & $0.59$ \tabularnewline
 & \multicolumn{1}{c|}{MAID ED  } &  & $1595\pm2\pm1$ & $131\pm3\pm1$ & $0.423\pm0.009\pm0.003$ & $(153\pm1\pm1)^{\circ}$ & $0.34$ 
\tabularnewline
\hline \hline
 $P_{13}(_{p}E_{1+})$  & SAID ED & $N(1720)\:3/2^{+}$ & $1651\pm7\pm2$ & $311\pm15\pm10$ & $0.108\pm0.001\pm0.008$ & 
$-(48\pm3\pm2)^{\circ}$ & $0.07$\tabularnewline
 & \multicolumn{1}{c|}{MAID ED  } &  & $1713\pm2\pm1$ & $239\pm4\pm3$ & $0.103\pm0.002\pm0.002$ & $-(21\pm1\pm1)^{\circ}$ & $0.55$ 
\tabularnewline
\hline \hline
$P_{13}(_{p}M_{1+})$  & SAID ED & $N(1720)\:3/2^{+}$ & $1637\pm3\pm14$ & $307\pm7\pm10$ & $0.071\pm0.002\pm0.002$ & 
$-(148\pm2\pm20)^{\circ}$ & $0.45$ \tabularnewline
 & MAID ED   &  & $1679\pm3\pm2$ & $243\pm6\pm4$ & $0.083\pm0.002\pm0.003$ & $-(63\pm1\pm2)^{\circ}$&$0.44$\tabularnewline
\hline \hline
$P_{31}(_{}M_{1-})$ & SAID ED & $\Delta(1910)\:1/2^{+}$ & $1778\pm16\pm4$ & $394\pm35\pm5$ & $0.356\pm0.037\pm0.016$ & 
$-(97\pm5\pm7)^{\circ}$ & $0.07$ \tabularnewline
 & MAID ED   &  & $1895\pm1\pm6$ & $326\pm2\pm1$ & $0.386\pm0.003\pm0.007$ & $(6\pm1\pm1)^{\circ}$& $0.93$\tabularnewline
\hline \hline
$D_{13}(_{p}M_{2-})$ & SAID ED & $N(1520)\:3/2^{-}$ & $1515\pm1\pm0$ & $110\pm2\pm1$ & $0.177\pm0.003\pm0.001$ & $(1\pm1\pm0)^{\circ}$ 
& $0.41$ \tabularnewline
 & MAID ED   &  & $1509\pm0.5\pm0.5$ & $102\pm1\pm1$ & $0.169\pm0.001\pm0.003$ & $(8\pm0.5\pm0.5)^{\circ}$&$0.46$\tabularnewline
\hline \hline
$D_{15}(_{p}E_{2+})$ & SAID ED  &$N(1675)\:5/2^{-}$  & $1657\pm3\pm2$ & $143\pm6\pm3$ & $0.012\pm0.002\pm0.002$ & 
$(66\pm3\pm2)^{\circ}$ & $0.33$\tabularnewline
 & MAID ED   &  & $1663\pm1\pm1$ & $137\pm2\pm1$ & $0.010\pm0.001\pm0.001$ & $(79\pm1\pm3)^{\circ}$&$0.25$\tabularnewline
\hline \hline
$D_{15}(_{p}M_{2+})$ & SAID ED  &$N(1675)\:5/2^{-}$ & $1656\pm2\pm3$ & $139\pm5\pm1$ & $0.028\pm0.002\pm0.001$ & 
$-(27\pm3\pm1)^{\circ}$ & $0.08$ \tabularnewline
 & MAID ED   &  & $1658\pm1\pm6$ & $138\pm1\pm1$ & $0.036\pm0.001\pm0.001$ & $-(20\pm1\pm0)^{\circ}$&$0.46$\tabularnewline
\hline
\hline
$D_{33}(_{}M_{2-})$  & SAID ED & $\Delta(1700)\:3/2^{-}$ & $1637\pm2\pm3$ & $273\pm5\pm1$ & $0.151\pm0.003\pm0.001$ & 
$-(16\pm1\pm3)^{\circ}$ & $0.92$ \tabularnewline
 & MAID ED   &  & $1645\pm1\pm3$ & $211\pm2\pm2$ & $0.125\pm0.001\pm0.002$ & $-(10\pm1\pm3)^{\circ}$& $0.93$\tabularnewline
\hline \hline
$F_{15}(_{p}M_{3-})$ & SAID ED & $N(1680)\:5/2^{+}$ & $1674\pm1\pm1$ & $112\pm3\pm2$ & $0.093\pm0.002\pm0.001$ & 
$-(14\pm3\pm2)^{\circ}$ & $0.63$ \tabularnewline
 & MAID ED   &  & $1642\pm1\pm10$ & $123\pm1\pm1$ & $0.112\pm0.001\pm0.001$ & $-(11\pm1\pm0)^{\circ}$& $0.96$\tabularnewline
\hline \hline
$F_{35}(_{}E_{3-})$ & SAID ED & $N(1905)\:5/2^{+}$ & $1817\pm5\pm2$ & $257\pm12\pm3$ & $0.049\pm0.003\pm0.001$ & $(6\pm3\pm2)^{\circ}$ 
& $0.05$ \tabularnewline
 & MAID ED   &  & $1842\pm4\pm8$ & $248\pm8\pm13$ & $0.017\pm0.002\pm0.002$ & $-(34\pm2\pm5)^{\circ}$&$0.09$\tabularnewline
\hline \hline
$F_{35}(_{}M_{3-})$ & SAID ED & $N(1905)\:5/2^{+}$ & $1815\pm4\pm2$ & $266\pm7\pm1$ & $0.046\pm0.002\pm0.001$ & $-(20\pm2\pm2)^{\circ}$ 
& $0.63$ \tabularnewline
 & MAID ED   &  & $1834\pm2\pm2$ & $288\pm4\pm5$ & $0.038\pm0.001\pm0.001$ & $-(27\pm1\pm2)^{\circ}$&$0.12$\tabularnewline
\hline \hline
$F_{37}(_{}E_{3+})$ & SAID ED & $\Delta(1950)\:7/2^{+}$ & $1879\pm3\pm2$ & $231\pm7\pm2$ & $0.014\pm0.001\pm0.001$ & 
$-(91\pm2\pm2)^{\circ}$ & $0.98$ \tabularnewline
 & MAID ED   &  & $1878\pm1\pm1$ & $222\pm3\pm3$ & $0.012\pm0.000\pm0.001$ & $-(115\pm2\pm2)^{\circ}$ & $0.18$ \tabularnewline
\hline \hline
\end{tabular}
\end{table*}

Together with the results in the previous tables, we now have a complete set of e.m. residues,
which allow us to relate the residues of the photoproduction multipoles to the normalized residues
$(NR)_{\gamma,\pi}^{h}$ and to the photo-decay amplitudes $A_h$ for helicity $h=1/2$ and $3/2\,$.

For consistency with the elastic and inelastic hadronic reactions, we first introduce the unitary
and dimensionless T-matrix elements $T_{\gamma,\pi}^{h}\,$.

Following the notation of Ref.~\cite{Workman:2013rca}, the $(\gamma,\pi)$ T-matrix element for
helicity $h$ is given by
\begin{equation}
T_{\gamma,\pi}^{h}=\sqrt{2\,k\,q}\; \mathcal{A}^{h}_\alpha\; C\,, \label{uniamp}
\end{equation}
where $\alpha$ denotes the partial wave and $k,q$ are the c.m. momenta of the photon and the pion.
The factor $C$ is $\sqrt{2/3}$ for isospin $3/2$ and $-\sqrt{3}$ for isospin $1/2$. The helicity
multipoles $\mathcal{A}^{h}_\alpha$ are given in terms of electric and magnetic multipoles
\begin{eqnarray}
\mathcal{A}_{\ell +}^{1/2} & = & -{1\over 2} \left[ (\ell +2) {E}_{\ell +} + \ell {M}_{\ell +}
\right] ,\label{helimult1}
\\
\mathcal{A}_{\ell +}^{3/2} & = & {1\over 2} \sqrt{\ell ( \ell + 2)} \left[ {E}_{\ell +} -{M}_{\ell
+} \right] , \label{helimult2}
\\
\mathcal{A}_{(\ell +1)-}^{1/2} & = & -{1\over 2} \left[\ell {E}_{(\ell +1)-} - (\ell +2 )
{M}_{(\ell +1)-} \right] , \label{helimult3}
\\
\mathcal{A}_{(\ell +1)-}^{3/2} & = & -{1\over 2} \sqrt{\ell ( \ell +2)} \left[ {E}_{(\ell +1)-} +
{M}_{(\ell +1) -} \right] , \label{helimult4}
\end{eqnarray}
with $J=\ell+1/2$ for '$+$' multipoles and $J=(\ell+1)-1/2$ for '$-$' multipoles, all having the
same total spin $J\,$.

Compared to the e.m. multipoles, which carry a dimension of length, the  T-matrix elements, used
here, are dimensionless and have residues defined by the pole term.
\begin{equation}
T_{\gamma,\pi}^{pole,h}(W)=\frac{R_{\gamma,\pi}^{h}}{M-W-i\Gamma/2}\,. \label{NRansatz}
\end{equation}

\begin{table*}[ht]
\caption{\label{tab:photo-decay} Normalized pion photoproduction T-matrix residues
(dimensionless) with helicity $1/2$ and $3/2$ and photo-decay amplitudes in units of $GeV^{-1/2}$.
The complex quantities are given in magnitudes and phases.}
\begin{tabular}{c|c|cccc|cccc}
\hline\hline
 Resonance & source & \multicolumn{2}{c}{$(NR)_{\gamma,\pi}^{1/2}$} & \multicolumn{2}{c|}{$(NR)_{\gamma,\pi}^{3/2}$} &
 \multicolumn{2}{c}{ $A_{1/2}$} & \multicolumn{2}{c}{ $A_{3/2}$} \\
\hline\hline
$\Delta(1232)\;3/2^+$& MAID ED & $ 0.0366(12)$ & $128(2)^\circ$ & $0.0722(20)$ & $141(2)^\circ$ & $0.130(2)$ & $161(2)^\circ$ & $0.258(3)$ 
& $174(2)^\circ$ \\
                     & SAID ED & $ 0.0363(11)$ & $134(2)^\circ$ & $0.0727(18)$ & $146(2)^\circ$ & $0.129(2)$ & $167(2)^\circ$ & $0.259(2)$ 
& $179(2)^\circ$ \\
\hline
$N(1440)\;1/2^+$     & MAID ED & $ 0.0165(11)$ & $126(4)^\circ$ & & & $0.058(1)$  & $-176(6)^\circ$ & &  \\
                     & SAID ED & $ 0.0156(23)$ & $109(8)^\circ$ & & & $0.055(3)$  &  $167(11)^\circ$ &&   \\
\hline
$N(1520)\;3/2^-$    &  MAID ED & $ 0.0068(7)$ & $175(6)^\circ$ & $0.0480(6)$ & $  5(1)^\circ$ & $0.019(2)$ & $-178(7)^\circ$ & $0.133(2)$ 
& $12(1)^\circ$ \\
                     & SAID ED & $ 0.0100(14)$ & $147(7)^\circ$ & $0.0482(25)$ & $  6(2)^\circ$ & $0.028(4)$ & $154(7)^\circ$ & $0.133(6)$ 
& $13(2)^\circ$ \\
\hline
$N(1535)\;1/2^-$     & MAID ED & $ 0.0276(20)$ &   $-6(10)^\circ$ & & & $ 0.071(3)$ &   $6(10)^\circ$ &&   \\
                     & SAID ED & $ 0.0289(49)$ &  $-29(9)^\circ$  & & & $ 0.074(10)$&  $-17(11)^\circ$ &&   \\
\hline
$\Delta(1620)\;1/2^-$& MAID ED & $ 0.0189(6)$ & $-32(2)^\circ$   & & & $0.065(1)$  & $19(2)^\circ$    &&   \\
                     & SAID ED & $ 0.0148(16)$ & $-47(6)^\circ$   & & & $0.051(1)$  & $ 4(9)^\circ$    &&   \\
\hline
$N(1650)\;1/2^-$     & MAID ED & $ 0.0496(38)$ &   $ 9(7)^\circ$  & & & $ 0.100(10)$  &  $46(6)^\circ$ &&   \\
                     & SAID ED & $ 0.0204(64)$ &  $-21(21)^\circ$ & & & $ 0.041(6)$   &  $16(27)^\circ$ &&   \\
\hline
$N(1675)\;5/2^-$    & MAID ED & $ 0.0038(3)$ & $  6(6)^\circ$ & $0.0055(3)$ & $-40(3)^\circ$  & $0.016(1)$ & $ 21(6)^\circ$ & $0.024(1)$ 
& $-25(4)^\circ$ \\
                     & SAID ED & $ 0.0036(5)$ & $ 10(12)^\circ$ &$0.0044(6)$ & $-55(7)^\circ$  & $0.015(2)$ & $ 25(12)^\circ$ & 
$0.019(2)$ & $-40(8)^\circ$ \\
\hline
$N(1680)\;5/2^+$    & MAID ED & $ 0.0096(7)$ & $150(5)^\circ$ & $0.0454(8)$ & $-10(1)^\circ$  & $0.027(2)$ & $156(5)^\circ$ & $0.129(2)$ 
& $-4(1)^\circ$ \\
                     & SAID ED & $ 0.0049(19)$ & $124(20)^\circ$ &$0.0433(17)$ & $-12(3)^\circ$  & $0.014(5)$ & $130(20)^\circ$ & 
$0.123(4)$ & $-6(3)^\circ$ \\
\hline
$\Delta(1700)\;3/2^-$& MAID ED & $ 0.0158(3)$ & $ -8(2)^\circ$ & $0.0166(5)$ & $ -2(3)^\circ$ & $0.125(2)$ & $ 20(2)^\circ$ & $0.132(4)$ 
& $ 27(3)^\circ$ \\
                     & SAID ED & $ 0.0141(7)$ & $-14(2)^\circ$ & $0.0117(12)$ & $ -1(3)^\circ$ & $0.112(3)$ & $ 15(3)^\circ$ & $0.093(7)$ 
& $ 28(5)^\circ$ \\
\hline
$N(1720)\;3/2^+$    & MAID ED & $ 0.0076(3)$ & $-41(2)^\circ$ & $0.0024(2)$ & $-159(4)^\circ$  & $0.069(1)$ & $ 17(3)^\circ$ & $0.022(3)$ 
& $-101(2)^\circ$ \\
                     & SAID ED & $ 0.0065(6)$ & $-72(7)^\circ$ &$0.0049(8)$ & $150(9)^\circ$    & $0.059(2)$ & $-14(8)^\circ$ & 
$0.045(5)$ & $-151(11)^\circ$ \\
\hline
$\Delta(1905)\;5/2^+$& MAID ED & $ 0.0019(2)$ & $ -32(5)^\circ$ & $0.0025(2)$ & $144(4)^\circ$ & $0.017(1)$ & $-10(6)^\circ$ & $0.023(1)$ 
& $166(5)^\circ$ \\
                     & SAID ED & $ 0.0017(2)$ & $-51(8)^\circ$ & $0.0042(2)$ & $167(3)^\circ$  & $0.015(2)$ & $-29(9)^\circ$ & $0.038(1)$ 
& $-172(4)^\circ$ \\
\hline
$\Delta(1910)\;1/2^+$& MAID ED & $ 0.0062(2)$ & $  -8(2)^\circ$  & & & $0.036(1)$  & $-80(2)^\circ$ & &  \\
                     & SAID ED & $ 0.0057(7)$ & $-111(13)^\circ$ & & & $0.033(5)$  &  $177(11)^\circ$ &&   \\
\hline
$\Delta(1950)\;7/2^+$& MAID ED & $ 0.0182(7)$ & $160(2)^\circ$ & $0.0240(9)$ & $165(2)^\circ$  & $0.090(2)$ & $-179(3)^\circ$ & 
$0.119(3)$ & $ -174(2)^\circ$ \\
                     & SAID ED & $ 0.0155(10)$ & $154(3)^\circ$ & $0.0193(14)$ & $161(3)^\circ$  & $0.076(4)$ & $175(4)^\circ$ & $0.095(5)$ 
& $-178(4)^\circ$ \\
\hline\hline
\end{tabular}
\end{table*}

In Table~\ref{tab:photo-decay} we list the normalized residues
\begin{equation}
(NR)_{\gamma,\pi}^h=\frac{R_{\gamma,\pi}^h}{\Gamma_p/2}\,,
\end{equation}
together with the photo-decay amplitudes
\begin{eqnarray}
A_h &=& C\,\sqrt{\frac{q_p}{k_p}\frac{2\pi(2J+1)W_p}{m_N Res_{\pi N}}}\,\mbox{Res}\,\mathcal{A}_\alpha^h\,,\\
    &=&\sqrt{\frac{\pi(2J+1)W_p}{k_p^2\, m_N Res_{\pi N}}}\; R_{\gamma,\pi}^{h}\,,
\end{eqnarray}
where the subscript $p$ denotes quantities evaluated at the pole position. For the elastic residues,
$Res_{\pi N}$, and the pole positions, $W_p=M_p-i\Gamma_p/2$, we use the values of the GWU/SAID
partial wave analysis, SP06~\cite{GWU}.

The errors shown for the normalized residues and for the photo-decay amplitudes are obtained by
error propagation from the uncertainties in the residues of the e.m. multipoles, listed in the
previous tables. We considered the total errors of the E,M residues, which arise from the fitting
and from the variation of the branch points. We also checked uncertainties from the pole positions,
however, these errors are significantly smaller than the residue errors and can be neglected.

In almost all cases, our results in Table~\ref{tab:photo-decay} show a very consistent behavior in
the comparison between the analyses of the MAID and SAID solutions, with deviations mostly within
$(1-2)\sigma$. An exception can be found for the $N(1650)1/2^-$. For this second resonance in the
$S_{11}$ partial wave, the normalized residues and photo-decay amplitudes in our analyses of MAID
and SAID differ by more than a factor of two. However, this is not too surprising, as a look in the
Particle Data Listings show that also the Breit-Wigner amplitudes differ by more than a factor of
two and even the elastic pole residues show very large deviations.
\\ \\ \noindent
Very recently, pole values for the photo-decay amplitudes of nucleon resonances were also analyzed
and published by the Bonn-Gatchina group~\cite{Bonn:2012}, the Argonne-Osaka
group~\cite{Kamano:2013} and the J\"ulich group~\cite{Ronchen:2014}. In Tables
\ref{tab:photo-decay-nucl} and \ref{tab:photo-decay-delt}, we show a comparison with our current
work. For many amplitudes, the magnitudes are in good agreement, while the residue phases differ
quite substantially. An exception is seen in the $\Delta(1232)$ resonance, only the J\"ulich
results are somewhat different. Similar to the elastic $\pi N$ residues, where the residue phase is
a measure of the non-resonant background, here also for the photoproduction residues we can assume
that the residue phases give a measure of the photoproduction background contributions. And this
part of the amplitude is less-well known, for most resonances, than the resonance contribution
itself. In the near future the efforts in the complete experiment analysis of pseudoscalar
photoproduction will certainly help to clarify this situation.

\begin{table*}[ht]
\caption{\label{tab:photo-decay-nucl} Comparison of pole positions and photo-decay amplitudes in
units of $GeV^{-1/2}$ between MAID (MD07), SAID (CM-12), J\"ulich (fit 2)~\cite{Ronchen:2014},
Bonn-Gatchina~\cite{Bonn:2012} and ANL-Osaka~\cite{Kamano:2013} for 4-star resonances with isospin
$1/2$. }
\begin{tabular}{|c|c|llll|llllll|}
\hline
 Resonance & source &\,& {$\mbox{Re}\,W_{p}$ }  &\;& {$-2\mbox{Im}\,W_{p}\;$ }  &\;&  \multicolumn{2}{c}{ $A_{1/2}$} &\;\;& 
\multicolumn{2}{c|}{ $A_{3/2}$} \\
\hline
$N(1440)\;1/2^+$     & MAID     && $1360(5)$ && $183(19)$ && $0.058(1)$ & $-176(6)^\circ$  &&  &  \\
                     & SAID     && $1367(2)$ && $190(5)$  && $0.055(3)$ & $167(11)^\circ$  &&  &  \\
                     & J\"ulich && $1353   $ && $212   $  && $0.054   $ & $ 137    ^\circ$ &&  &  \\
                     & BnGa     && $1370(4)$ && $190(7)$  && $0.044(7)$ & $  142(5)^\circ$ &&  &  \\
                     & ANL-O    && $1374   $ && $152   $  && $0.050   $ & $  -12   ^\circ$ &&  &  \\
\hline
$N(1520)\;3/2^-$     & MAID     && $1509(1)$ && $104(8)$ && $0.019(2)$ & $-178(7)^\circ$ && $0.133(2)$ & $ 12(1)^\circ$ \\
                     & SAID     && $1514(2)$ && $110(5)$ && $0.028(4)$ & $ 154(7)^\circ$ && $0.133(6)$ & $ 13(2)^\circ$ \\
                     & J\"ulich && $1519   $ && $110   $ && $0.024   $ & $ 156   ^\circ$ && $0.117   $ & $19    ^\circ$ \\
                     & BnGa     && $1507(3)$ && $111(5)$ && $0.021(4)$ & $180(5) ^\circ$ && $0.132(9)$ & $  2(4)^\circ$ \\
                     & ANL-O    && $1501   $ && $ 78   $ && $0.038   $ & $  3    ^\circ$ && $0.094   $ & $-173  ^\circ$ \\
\hline
$N(1535)\;1/2^-$     & MAID     && $1516(3)$ && $ 94(5)$ && $ 0.071(3)$ &   $6(10)^\circ$  &&  &  \\
                     & SAID     && $1501(6)$ && $95(11)$ && $ 0.074(10)$&  $-17(11)^\circ$  &&  &  \\
                     & J\"ulich && $1498   $ && $ 74   $  && $0.050   $ & $ -45    ^\circ$ &&  &  \\
                     & BnGa     && $1501(4)$ && $134(11)$ && $0.116(10)$& $    7(6)^\circ$ &&  &  \\
                     & ANL-O    && $1482   $ && $196   $  && $0.161   $ & $    9   ^\circ$ &&  &  \\
\hline
$N(1650)\;1/2^-$     & MAID     && $1678(4)$ && $135(5)$ && $ 0.100(10)$  &  $46(6)^\circ$  &&  &  \\
                     & SAID     && $1655(11)$ && $127(17)$ && $ 0.041(6)$   &  $16(27)^\circ$  &&  &  \\
                     & J\"ulich && $1677   $ && $146   $  && $0.023   $ & $ -29    ^\circ$ &&  &  \\
                     & BnGa     && $1647(6)$ && $103(8)$ && $0.033(7)$& $  -9 (15)^\circ$ &&  &  \\
                     & ANL-O    && $1656   $ && $170   $  && $0.040   $ & $  -44   ^\circ$ &&  &  \\
\hline
$N(1675)\;5/2^-$     & MAID     && $1661(10)$ && $138(4)$ && $0.016(1)$ & $ 21(6)^\circ$ && $0.024(1)$ & $-25(4)^\circ$ \\
                     & SAID     && $1657(6)$ && $141(11)$ && $0.015(2)$ & $ 25(12)^\circ$ && $0.019(2)$ & $-40(8)^\circ$ \\
                     & J\"ulich && $1650   $ && $126   $ && $0.022   $ & $  38   ^\circ$ && $0.036   $ & $-41    ^\circ$ \\
                     & BnGa     && $1654(4)$ && $151(5)$ && $0.024(3)$ & $-16(5) ^\circ$ && $0.026(8)$ & $-19(6)^\circ$ \\
                     & ANL-O    && $1650   $ && $150   $ && $0.005   $ & $-22    ^\circ$ && $0.033   $ & $-23  ^\circ$ \\
\hline
$N(1680)\;5/2^+$     & MAID     && $1653(22)$ && $121(6)$ && $0.027(2)$ & $156(5)^\circ$ && $0.129(2)$ & $-4(1)^\circ$ \\
                     & SAID     && $1674(3)$ && $113(6)$ && $0.014(5)$ & $130(20)^\circ$ && $0.123(4)$ & $-6(3)^\circ$ \\
                     & J\"ulich && $1666   $ && $108   $ && $0.013   $ & $ 120   ^\circ$ && $0.126   $ & $-24    ^\circ$ \\
                     & BnGa     && $1676(6)$ && $113(4)$ && $0.013(4)$ & $155(22) ^\circ$ && $0.134(5)$ & $-2(4)^\circ$ \\
                     & ANL-O    && $1665   $ && $ 98   $ && $0.053   $ & $-5     ^\circ$ && $0.038   $ & $-177  ^\circ$ \\
\hline
$N(1720)\;3/2^+$     & MAID     && $1696(22)$ && $241(12)$ && $0.069(1)$ & $ 17(3)^\circ$ && $0.022(3)$ & $-101(2)^\circ$ \\
                     & SAID     && $1644(24)$&& $309(27)$ && $0.059(2)$ & $-14(8)^\circ$ && $0.045(5)$ & $-151(11)^\circ$ \\
                     & J\"ulich && $1717   $ && $208   $ && $0.051   $ & $ -8   ^\circ$ && $0.014   $ & $ 37    ^\circ$ \\
                     & BnGa     && $1660(30)$ && $450(100)$ && $0.110(45)$ & $ 0(40) ^\circ$ && $0.150(35)$ & $65(35)^\circ$ \\
                     & ANL-O    && $1703   $ && $ 140   $ && $0.234   $ & $ 2     ^\circ$ && $0.070   $ & $173  ^\circ$ \\
\hline
\end{tabular}
\end{table*}

\begin{table*}[ht]
\caption{\label{tab:photo-decay-delt} Same as in Table~\ref{tab:photo-decay-nucl} for 4-star
resonances with isospin $3/2$.}
\begin{tabular}{|c|c|llll|llllll|}
\hline
 Resonance & source &\,& {$\mbox{Re}\,W_{p}$ }  &\;& {$-2\mbox{Im}\,W_{p}\;$ }  &\;&  \multicolumn{2}{c}{ $A_{1/2}$} &\;\;& 
\multicolumn{2}{c|}{ $A_{3/2}$} \\
\hline
$\Delta(1232)\;3/2^+$& MAID     && $1211(1)$ && $ 99(1)$ && $0.130(2)$ & $161(2)^\circ$ && $0.258(3)$ & $174(2)^\circ$ \\
                     & SAID     && $1211(1)$ && $101(1)$ && $0.129(2)$ & $167(2)^\circ$ && $0.259(2)$ & $179(2)^\circ$ \\
                     & J\"ulich && $1220   $ && $ 86   $ && $0.114   $ & $   153^\circ$ && $0.229   $ & $   165^\circ$ \\
                     & BnGa     && $1210(1)$ && $ 99(2)$ && $0.131(4)$ & $161(2)^\circ$ && $0.254(5)$ & $171(1)^\circ$ \\
                     & ANL-O    && $1211   $ && $102   $ && $0.133   $ & $   165^\circ$ && $0.257   $ & $177   ^\circ$ \\
\hline
$\Delta(1620)\;1/2^-$& MAID     && $1595(3)$ && $131(4)$ && $0.065(1)$  & $19(2)^\circ$  &&  &  \\
                     & SAID     && $1596(4)$ && $124(7)$ && $0.051(1)$  & $ 4(9)^\circ$  &&  &  \\
                     & J\"ulich && $1599   $ && $ 71   $  && $0.028   $ & $-88     ^\circ$ &&  &  \\
                     & BnGa     && $1597(4)$ && $130(9)$ && $0.052(5)$& $  -9(9)^\circ$ &&  &  \\
                     & ANL-O    && $1592   $ && $136   $  && $0.113   $ & $  -1   ^\circ$ &&  &  \\
\hline
$\Delta(1700)\;3/2^-$& MAID     && $1647(6)$ && $217(13)$ && $0.125(2)$ & $ 20(2)^\circ$ && $0.132(4)$ & $ 27(3)^\circ$ \\
                     & SAID     && $1644(12)$&& $264(20)$ && $0.112(3)$ & $ 15(3)^\circ$ && $0.093(7)$ & $ 28(5)^\circ$ \\
                     & J\"ulich && $1675   $ && $303   $ && $0.109   $ & $ -12   ^\circ$ && $0.111   $ & $ 21    ^\circ$ \\
                     & BnGa     && $1680(10)$ && $305(15)$ && $0.170(20)$ & $ 50(15) ^\circ$ && $0.170(25)$ & $45(10)^\circ$ \\
                     & ANL-O    && $1707   $ && $ 340   $ && $0.059   $ & $-70     ^\circ$ && $0.125   $ & $-75  ^\circ$ \\
\hline
$\Delta(1905)\;5/2^+$& MAID     && $1838(16)$ && $268(41)$ && $0.017(1)$ & $-10(6)^\circ$ && $0.023(1)$ & $166(5)^\circ$ \\
                     & SAID     && $1816(8)$&& $262(17)$ && $0.015(2)$ & $-29(9)^\circ$ && $0.038(1)$ & $-172(4)^\circ$ \\
                     & J\"ulich && $1770   $ && $259   $ && $0.013   $ & $ 19   ^\circ$ && $0.072   $ & $ 67    ^\circ$ \\
                     & BnGa     && $1805(10)$ && $300(15)$ && $0.025(5)$ & $-23(15) ^\circ$ && $0.050(4)$ & $180(10)^\circ$ \\
                     & ANL-O    && $1765   $ && $ 188   $ && $0.008   $ & $-97     ^\circ$ && $0.018   $ & $-90  ^\circ$ \\
\hline
$\Delta(1910)\;1/2^+$& MAID     && $1895(7)$ && $326(3)$   && $0.036(1)$  & $-80(2)^\circ$    &&  &  \\
                     & SAID     && $1778(20)$ && $394(40)$ && $0.033(5)$  &  $177(11)^\circ$  &&  &  \\
                     & J\"ulich && $1788   $ && $575   $  && $0.246   $ & $ -133    ^\circ$ &&  &  \\
                     & BnGa     && $1850(40)$ && $350(45)$ && $0.023(9)$& $  40 (90)^\circ$ &&  &  \\
                     & ANL-O    && $1854   $ && $368   $  && $0.052   $ & $  170   ^\circ$ &&  &  \\
\hline
$\Delta(1950)\;7/2^+$& MAID     && $1888(12)$ && $247(31)$ && $0.090(2)$ & $-179(3)^\circ$ && $0.118(3)$ & $-174(2)^\circ$ \\
                     & SAID     && $1882(8)$  && $231(9)$ && $0.076(4)$ & $175(4)^\circ$ && $0.095(5)$ & $ -178(4)^\circ$ \\
                     & J\"ulich && $1884   $ && $234   $ && $0.071   $ & $ 151   ^\circ$ && $0.089   $ & $ 155    ^\circ$ \\
                     & BnGa     && $1890(4)$ && $243(8)$ && $0.072(4)$ & $ 173(5) ^\circ$ && $0.096(5)$ & $173(5)^\circ$ \\
                     & ANL-O    && $1872   $ && $ 206   $ && $0.062   $ & $ 171   ^\circ$ && $0.076   $ & $-178  ^\circ$ \\
\hline
\end{tabular}
\end{table*}

\section{Summary and Conclusions}

In this work, we have applied the L+P method to the partial wave amplitudes of the MAID and SAID
solutions for single-pion photoproduction. We have analyzed both energy-dependent and single-energy
solutions, and have determined pole positions and residues from electromagnetic multipoles in the
region up to $W\sim 2$~GeV. The pole positions are compared to values listed in the Particle Data
Tables, and show almost perfect agreement with data coming from other channels. Presently,
inelastic residues are very sparse in the Particle Data Tables, with no residues for meson
photoproduction yet listed. However, since the PDG is recommending the replacement of Breit-Wigner
parameters by pole parameters in future listings, we find that the L+P method, being controllably
model-dependent, is a good method to extract this information from both ED and SE partial wave
amplitudes.
\\ \\ \noindent
We have found that, for all partial waves, the first resonance state can be consistently analyzed
with the L+P technique and good agreement on the pole positions can be observed. This also gives us
confidence in the determination of the complex residues for pion photoproduction. In the special
case of the $S_{11}$ partial wave, the second (4-star) resonance, $N(1650)1/2^-$ can be equally
well analyzed, while a third $N(1895)1/2^-$ can only be found in the MAID ED solution. For most
other (2- and 3-star) resonances, our analysis finds large deviations among the four different
solutions. From these resonances, the $\Delta(1600)3/2^+$ is best determined, other states as
$N(1710)1/2^+$, $N(1700)3/2^-$ and $N(2000)5/2^+$ give inconclusive results. All three of them,
however, can alternatively be replaced by a complex branch point in the appropriate $P_{11},
D_{13}$ and $F_{15}$ partial waves. For these partial waves, we have demonstrated that the
amplitudes can be similarly described by either a real branch point and two resonances or a complex
branch point and only one resonance.
\\ \\ \noindent
Furthermore, for all partial waves, we have investigated the four-star resonances in the energy
region $W<2$~GeV with respect to the normalized T-matrix residues and the photo-decay amplitudes at
the pole positions. We have compared our results with other very recently published analyses and
find good agreement for dominant amplitudes, but also considerable deviations for smaller
amplitudes or amplitudes of nucleon resonances that are less well determined.
\\ \\ \noindent
In conclusion, we have found that a single-channel partial wave analysis can consistently determine
the pole position and parameters of the lowest nucleon resonances, but cannot distinguish between
higher resonances and alternative complex branch points. However, with the additional information
of other decay channels, especially with three-body final states, a unique determination should be
possible.
\\ \\ \noindent

\begin{acknowledgments}
This work was supported in part by the U.S. Department of Energy Grant DE-FG02-99ER41110, the
Deutsche Forschungsgemeinschaft (SFB 1044).
\end{acknowledgments}

\begin{table*}[t]

\centerline{\textbf{APPENDIX \vspace*{1.5cm}}}

 Tables VIII and IX compare results with the third
branch point either fixed, based on the threshold for a dominant inelastic channel, or allowed to
adjust for a best fit. The variation is used in the estimation of systematic errors, as discussed
in Section III.B.

\caption{\label{tab:paramED} Parameters from L+P expansion are given for GWU/SAID and MAID energy
dependent (ED) solutions. $N_r$ is the number of resonance poles, $x_P, x_Q, x_R$ are branch points
in $MeV$.}
\begin{tabular}{c|crclc|crclc}
\hline \hline
\multirow{3}{*}{Multipole} & \multicolumn{10}{c}{Source}\tabularnewline
\cline{2-11}
 & \multicolumn{5}{c|}{SAID  ED} & \multicolumn{5}{c}{MAID ED}\tabularnewline
\cline{2-11}
 & $N_{r}$ & $x_{P}$ & $x_{Q}$ & $x_{R}$ & $10^2 D_{dp}$ & $N_{r}$ & $x_{P}$ & $x_{Q}$ & $x_{R}$ & $10^2 D_{dp}$\tabularnewline
\hline \hline
\multirow{3}{*}{$S_{11}(_{p}E_{0+})$} & $2$ & $142$ & $1077^{\pi N}$ & $1215^{\pi\pi N}$ & $0.49$ & $3$ & $-3778$ & $1077^{\pi N}$ & 
$1215^{\pi\pi N}$ & $1.20$\tabularnewline
 & $2$ & $900$ & $1077^{\pi N}$ & $1486^{\eta N}$ & $0.35$ & $3$ & $-131$ & $1077^{\pi N}$ & $1486^{\eta N}$ & $1.01$\tabularnewline
 & $2$ & $889$ & $1077^{\pi N}$ & $1495^{free}$ & $0.31$ & $3$ & $-393$ & $1077^{\pi N}$ & $1379^{free}$ & $0.98$\tabularnewline
\hline\hline
\multirow{3}{*}{$P_{11}(_{p}M_{1-})$} & $2$ & $900$ & $1077^{\pi N}$ & $1215^{\pi\pi N}$ & $0.13$ & $2$ & $309$ & $1077^{\pi N}$ & 
$1215^{\pi\pi N}$ & $0.31$\tabularnewline
 & $2$ & $900$ & $1077^{\pi N}$ & $1370^{Real(\pi\Delta)}$ & $0.15$ & $2$ & $494$ & $1077^{\pi N}$ & $1370^{Real(\pi\Delta)}$ & 
$0.28$\tabularnewline
 & $2$ & $871$ & $1077^{\pi N}$ & $1375{}^{free}$ & $0.12$ & $2$ & $123$ & $1077^{\pi N}$ & $1515^{free}$ & $0.15$\tabularnewline
\hline \hline
\multirow{3}{*}{$P_{33}(_{p}E_{1+})$} & $2$ & $883$ & $1077^{\pi N}$ & $1215^{\pi\pi N}$ & $0.12$ & $2$ & $838$ & $1077^{\pi N}$ & 
$1215^{\pi\pi N}$ & $0.20$\tabularnewline
 & $2$ & $899$ & $1077^{\pi N}$ & $1370^{Real(\pi\Delta)}$ & $0.12$ & $2$ & $167$ & $1077^{\pi N}$ & $1370^{Real(\pi\Delta)}$ & 
$0.19$\tabularnewline
 & $2$ & $818$ & $1077^{\pi N}$ & $1218^{free}$ & $0.09$ & $2$ & $534$ & $1077^{\pi N}$ & $1222^{free}$ & $0.09$\tabularnewline
\hline \hline
\multirow{3}{*}{$P_{33}(M_{1+})$} & $2$ & $900$ & $1077^{\pi N}$ & $1215^{\pi\pi N}$ & $0.08$ & $2$ & $-5238$ & $1077^{\pi N}$ & 
$1215^{\pi\pi N}$ & $1.29$\tabularnewline
 & $2$ & $788$ & $1077^{\pi N}$ & $1370^{Real(\pi\Delta)}$ & $0.05$ & $2$ & $-218$ & $1077^{\pi N}$ & $1370^{Real(\pi\Delta)}$ & 
$1.50$\tabularnewline
 & $2$ & $507$ & $1077^{\pi N}$ & $1238^{free}$ & $0.02$ & $2$ & $900$ & $1077^{\pi N}$ & $1265^{free}$ & $0.89$\tabularnewline
\hline \hline
\multirow{4}{*}{$D_{13}(_{p}E_{2-})$} & $2$ & $900$ & $1077^{\pi N}$ & $1215^{\pi\pi N}$ & $0.28$ & $1$ & $485$ & $1077^{\pi N}$ & 
$1215^{\pi\pi N}$ & $0.35$\tabularnewline
 & $2$ & $832$ & $1077^{\pi N}$ & $1370^{Real(\pi\Delta)}$ & $0.34$ & $1$ & $528$ & $1077^{\pi N}$ & $1370^{Real(\pi\Delta)}$ & 
$0.33$\tabularnewline
 & $2$ & $900$ & $1077^{\pi N}$ & $1700^{Real(\rho N)}$ & $0.36$ & $1$ & $48$ & $1077^{\pi N}$ & $1700^{Real(\rho N)}$ & 
$0.17$\tabularnewline
 & $2$ & $900$ & $1077^{\pi N}$ & $1232^{free}$ & $0.28$ & $1$ & $898$ & $1077^{\pi N}$ & $1717^{free}$ & $0.06$\tabularnewline
\hline \hline
\multirow{3}{*}{$D_{33}(_{p}E_{2-})$} & $1$ & $900$ & $1077^{\pi N}$ & $1215^{\pi\pi N}$ & $0.18$ & $1$ & $900$ & $1077^{\pi N}$ & 
$1215^{\pi\pi N}$ & $0.45$\tabularnewline
 & $1$ & $537$ & $1077^{\pi N}$ & $1370^{Real(\pi\Delta)}$ & $0.19$ & $1$ & $609$ & $1077^{\pi N}$ & $1370^{Real(\pi\Delta)}$ & 
$0.46$\tabularnewline
 & $1$ & $900$ & $1077^{\pi N}$ & $1105^{free}$ & $0.15$ & $1$ & $900$ & $1077^{\pi N}$ & $1222^{free}$ & $0.34$\tabularnewline
\hline\hline
\multirow{4}{*}{$F_{15}(_{p}E_{3-})$} & $1$ & $900$ & $1077^{\pi N}$ & $1215^{\pi\pi N}$ & $0.01$ & $2$ & $899$ & $1077^{\pi N}$ & 
$1215^{\pi\pi N}$ & $0.11$\tabularnewline
 & $1$ & $856$ & $1077^{\pi N}$ & $1370^{Real(\pi\Delta)}$ & $0.02$ & $2$ & $883$ & $1077^{\pi N}$ & $1370^{Real(\pi\Delta)}$ & 
$0.14$\tabularnewline
 & $1$ & $898$ & $1077^{\pi N}$ & $1700^{Real(\rho N)}$ & $0.02$ & $2$ & $898$ & $1077^{\pi N}$ & $1700^{Real(\rho N)}$ & 
$0.12$\tabularnewline
 & $1$ & $900$ & $1077^{\pi N}$ & $1222{}^{free}$ & $0.01$ & $2$ & $-118$ & $1077^{\pi N}$ & $1126^{free}$ & $0.10$\tabularnewline
\hline\hline
\multirow{4}{*}{$F_{37}(_{p}E_{3+})$}
 & $1$ & $897$ & $1077^{\pi N}$ & $1370^{Real(\pi\Delta)}$ & $0.01$ & $1$ & $-1931$ & $1077^{\pi N}$ & $1370^{Real(\pi\Delta)}$ & 
$0.03$\tabularnewline
 & $1$ & $755$ & $1077^{\pi N}$ & $1700^{Real(\rho N)}$ & $0.01$ & $1$ & $-977$ & $1077^{\pi N}$ & $1700^{Real(\rho N)}$ & 
$0.03$\tabularnewline
 & $1$ & $900$ & $1077^{\pi N}$ & $1285^{free}$ & $0.01$ & $1$ & $-215$ & $1077^{\pi N}$ & $1230^{free}$ & $0.02$\tabularnewline
\hline \hline
\end{tabular}
\end{table*}

\begin{table*}[ht]
\caption{\label{tab:paramSES} Parameters from  L+P expansion are given for GWU/SAID and MAID single
energy (SE) solutions. $N_r$ is the number of resonance poles, $x_P, x_Q, x_R$ are branch points in
$MeV$.}
\begin{tabular}{c|crclc|crclc}
\hline \hline
\multirow{3}{*}{Multipole} & \multicolumn{10}{c}{Source}\tabularnewline
\cline{2-11}
 & \multicolumn{5}{c|}{MAID SE} & \multicolumn{5}{c}{SAID SE}\tabularnewline
\cline{2-11}
 & $N_{r}$ & $x_{P}$ & $x_{Q}$ & $x_{R}$ & $\chi^{2}_{dp}$ & $N_{r}$ & $x_{P}$ & $x_{Q}$ & $x_{R}$ & $\chi^{2}_{dp}$\tabularnewline
\hline \hline
\multirow{3}{*}{$S_{11}(_{p}E_{0+})$} & $2$ & $-950$ & $1077^{\pi N}$ & $1215^{\pi\pi N}$ & $3.88$ & $2$ & $-9935$ & $1077^{\pi N}$ & 
$1215^{\pi\pi N}$ & $3.02$\tabularnewline
 & $2$ & $-395$ & $1077^{\pi N}$ & $1486^{\eta N}$ & $3.86$ & $2$ & $-4986$ & $1077^{\pi N}$ & $1486^{\eta N}$ & $2.53$\tabularnewline
 & $2$ & $876$ & $1077^{\pi N}$ & $1491^{free}$ & $3.53$ & $2$ & $556$ & $1077^{\pi N}$ & $1499{}^{free}$ & $2.47$\tabularnewline
\hline\hline
\multirow{3}{*}{$P_{11}(_{p}M_{1-})$} & $2$ & $-1037$ & $1077^{\pi N}$ & $1215^{\pi\pi N}$ & $3.03$ & $1$ & $-1256$ & $1077^{\pi N}$ & 
$1215^{\pi\pi N}$ & $3.02$\tabularnewline
 & $2$ & $-810$ & $1077^{\pi N}$ & $1370^{Real(\pi\Delta)}$ & $2.74$ & $1$ & $-12191$ & $1077^{\pi N}$ & $1370^{Real(\pi\Delta)}$ & 
$3.05$\tabularnewline
 & $2$ & $-3417$ & $1077^{\pi N}$ & $1362^{free}$ & $2.73$ & $1$ & $-10673$ & $1077^{\pi N}$ & $1324^{free}$ & $2.97$\tabularnewline
\hline\hline
\multirow{3}{*}{$P_{33}(_{p}E_{1+})$} & $1$ & $-2754$ & $1077^{\pi N}$ & $1215^{\pi\pi N}$ & $3.38$ & $1$ & $754$ & $1077^{\pi N}$ & 
$1215^{\pi\pi N}$ & $3.00$\tabularnewline
 & $1$ & $-1759$ & $1077^{\pi N}$ & $1370^{Real(\pi\Delta)}$ & $3.34$ & $1$ & $615$ & $1077^{\pi N}$ & $1370^{Real(\pi\Delta)}$ & 
$3.02$\tabularnewline
 & $1$ & $46$ & $1077^{\pi N}$ & $1467{}^{free}$ & $3.21$ & $1$ & $-1267$ & $1077^{\pi N}$ & $1155{}^{free}$ & $2.98$\tabularnewline
\hline \hline
\multirow{3}{*}{$P_{33}(M_{1+})$} & $1$ & $-1670$ & $1077^{\pi N}$ & $1215^{\pi\pi N}$ & $3.26$ & $2$ & $-1116$ & $1077^{\pi N}$ & 
$1215^{\pi\pi N}$ & $2.94$\tabularnewline
 & $1$ & $-7265$ & $1077^{\pi N}$ & $1370^{Real(\pi\Delta)}$ & $3.27$ & $2$ & $60$ & $1077^{\pi N}$ & $1370^{Real(\pi\Delta)}$ & 
$2.86$\tabularnewline
 & $1$ & $-387$ & $1077^{\pi N}$ & $1250^{free}$ & $3.24$ & $2$ & $639$ & $1077^{\pi N}$ & $1236^{free}$ & $2.84$\tabularnewline
\hline \hline
\multirow{4}{*}{$D_{13}(_{p}E_{2-})$} & $1$ & $-6892$ & $1077^{\pi N}$ & $1215^{\pi\pi N}$ & $2.77$ & $2$ & $775$ & $1077^{\pi N}$ & 
$1215^{\pi\pi N}$ & $2.68$\tabularnewline
 & $1$ & $-96$ & $1077^{\pi N}$ & $1370^{Real(\pi\Delta)}$ & $2.79$ & $2$ & $716$ & $1077^{\pi N}$ & $1370^{Real(\pi\Delta)}$ & 
$2.66$\tabularnewline
 & $1$ & $-232$ & $1077^{\pi N}$ & $1700^{Real(\rho N)}$ & $3.02$ & $2$ & $831$ & $1077^{\pi N}$ & $1700^{Real(\rho N)}$ & 
$2.57$\tabularnewline
 & $1$ & $900$ & $1077^{\pi N}$ & $1193^{free}$ & $2.61$ & $2$ & $179$ & $1077^{\pi N}$ & $1737^{free}$ & $2.33$\tabularnewline
\hline \hline
\multirow{3}{*}{$D_{33}(_{p}E_{2-})$} & $1$ & $862$ & $1077^{\pi N}$ & $1215^{\pi\pi N}$ & $5.04$ & $1$ & $638$ & $1077^{\pi N}$ & 
$1215^{\pi\pi N}$ & $2.53$\tabularnewline
 & $1$ & $835$ & $1077^{\pi N}$ & $1370^{Real(\pi\Delta)}$ & $4.71$ & $1$ & $837$ & $1077^{\pi N}$ & $1370^{Real(\pi\Delta)}$ & 
$2.45$\tabularnewline
 & $1$ & $899$ & $1077^{\pi N}$ & $1556{}^{free}$ & $4.61$ & $1$ & $-663$ & $1077^{\pi N}$ & $1374{}^{free}$ & $2.36$\tabularnewline
\hline \hline
\multirow{4}{*}{$F_{15}(_{p}E_{3-})$} & $1$ & $-27038$ & $1077^{\pi N}$ & $1215^{\pi\pi N}$ & $3.63$ & $2$ & $900$ & $1077^{\pi N}$ & 
$1215^{\pi\pi N}$ & $3.34$\tabularnewline
 & $1$ & $-923$ & $1077^{\pi N}$ & $1370^{Real(\pi\Delta)}$ & $3.93$ & $2$ & $-139$ & $1077^{\pi N}$ & $1370^{Real(\pi\Delta)}$ & 
$3.26$\tabularnewline
 & $1$ & $-715$ & $1077^{\pi N}$ & $1700^{Real(\rho N)}$ & $3.08$ & $2$ & $900$ & $1077^{\pi N}$ & $1700^{Real(\rho N)}$ & 
$2.86$\tabularnewline
 & $1$ & $90$ & $1077^{\pi N}$ & $1705^{free}$ & $3.03$ & $2$ & $-165$ & $1077^{\pi N}$ & $1717{}^{free}$ & $2.69$\tabularnewline
\hline \hline
\multirow{4}{*}{$F_{37}(_{p}E_{3+})$}
 & $1$ & $894$ & $1077^{\pi N}$ & $1370^{Real(\pi\Delta)}$ & $2.49$ & $1$ & $-146$ & $1077^{\pi N}$ & $1370^{Real(\pi\Delta)}$ & 
$1.66$\tabularnewline
 & $1$ & $70$ & $1077^{\pi N}$ & $1700^{Real(\rho N)}$ & $2.06$ & $1$ & $667$ & $1077^{\pi N}$ & $1700^{Real(\rho N)}$ & 
$1.68$\tabularnewline
 & $1$ & $-247$ & $1077^{\pi N}$ & $1649^{free}$ & $1.87$ & $1$ & $-1504$ & $1077^{\pi N}$ & $1270^{free}$ & $1.64$\tabularnewline
\hline  \hline
\end{tabular}
\end{table*}


\clearpage

\begin{thebibliography}{9}

\bibitem{PDG}
        J. Beringer et al. (Particle Data Group),
        Phys. Rev. \textbf{D86}, 010001 (2012).
\bibitem{Camogli2012}
        International Workshop on New partial wave analysis tools for next generation hadron
        spectroscopy experiments, June 20-22, 2012, Camogli, Italy.
        [http://www.ge.infn.it/$\sim$athos12].
\bibitem{Kloster2013}
        ATHOS 2013-International Workshop on New Partial-Wave Analysis Tools for Next-Generation Hadron
        Spectroscopy Experiments, 21-24 May 2013 Kloster Seeon, Germany,
        [http://intern.universe-cluster.de/indico/event/2857].
\bibitem{Doering} M. D\"{o}ring, C. Hanhart, F. Huang, S. Krewald, and U.-G. Mei\ss{}ner, Nucl. Phys. \textbf{A829}, 170 (2009), and 
references therein.
\bibitem{EBAC} B. Juli\'{a}-D\'{i}az, H. Kamano, T.-S. H. Lee, A. Matsuyama, T. Sato, N. Suzuki, Phys. Rev. \textbf{C80}, 025207 (2009), 
and references therein.
\bibitem{CMB} R. E. Cutkosky, C. P. Forsyth, R. E. Hendrick, and R. L. Kelly, Phys. Rev. \textbf{D 20}, 2839 (1979).
\bibitem{Zagreb} M. Batini\'{c}, I. \v{S}laus, A. \v{S}varc, and B. M. K. Nefkens, Phys. Rev. \textbf{C 51}, 2310 (1995); M. 
Batini\'{c} et al., Phys. Scr. \textbf{58}, 15 (1998).
\bibitem{Bonn:2012} A. V. Anisovich, R. Beck, E. Klempt, V. A. Nikonov, A. V. Sarantsev, U. Thoma,  Eur. Phys. J. \textbf{A48}, 15 
(2012); [http://pwa.hiskp.uni-bonn.de/], and references therein.
\bibitem{Hoehler93} G. H\"{o}hler, $\pi$N Newsletter \textbf{9}, 1 (1993).
\bibitem{Kelkar} N.G. Kelkar, M. Nowakowski,  Phys. Rev. \textbf{A78}, 012709 (2008), and references therein.
\bibitem{ChewMandelstam} G. F. Chew and S. Mandelstam, Phys. Rev. \textbf{119}, 467 (1960).
\bibitem{Ceci2008} S. Ceci, J. Stahov, A. \v{S}varc, S. Watson, and B. Zauner, Phys. Rev. \textbf{D 77}, 116007 (2008).
\bibitem{Svarc2013}
        A. \v{S}varc, M. Had\v{z}imehmedovi\'{c}, H. Osmanovi\'{c}, J. Stahov, L. Tiator, R. L. Workman,
        Phys. Rev. \textbf{C 88}, 035206 (2013).
\bibitem{GWU}
        R. A. Arndt, W. J. Briscoe, I. I. Strakovsky, and R. L. Workman,
        Phys. Rev. \textbf{C 74}, 045205 (2006);         [http://gwdac.phys.gwu.edu/analysis/]
\bibitem{GWU1} R. L. Workman, R. A. Arndt, W. J. Briscoe, M. W. Paris, and I. I. Strakovsky,  Phys. Rev. \textbf{C 86}, 035202 (2012).
\bibitem{DMT0}G. Y. Chen, S. S. Kamalov, S. N. Yang, D. Drechsel, and L. Tiator, Phys. Rev. \textbf{C 76}, 035206 (2007).
\bibitem{DMT} L. Tiator, S. S. Kamalov, S. Ceci, Guan Yeu Chen, D. Drechsel, A. Svarc, and Shin Nan Yang, Phys. Rev.  \textbf{C 82},
055203 (2010).
\bibitem{ses}
        R. L. Workman, M. W. Paris, W. J. Briscoe, L. Tiator, S. Schumann, M. Ostrick, S. S. Kamalov,
        Eur. Phys. J. \textbf{A47}, 143 (2011).
\bibitem{bnga_other}
        A.~V.~Anisovich, R.~Beck, E.~Klempt, V.~A.~Nikonov, A.~V.~Sarantsev, U.~Thoma and Y.~Wunderlich,
          Eur.\ Phys.\ J.\ A {\bf 49}, 121 (2013).
\bibitem{Mittag-Leffler}
        M. Hazewinkel:
        \emph{Encyclopaedia of Mathematics}, Vol. 6,  Springer, 1990, p. 251.
\bibitem{Ciulli}
        S. Ciulli and J. Fischer, Nucl. Phys. 24, 465 (1961).
\bibitem{Ciulli1}
        I. Ciulli, S. Ciulli, and J. Fischer, Nuovo Cimento \textbf{23}, 1129 (1962).
\bibitem{Pietarinen}
        E. Pietarinen, Nuovo Cimento \textbf{12A}, 522 (1972).
\bibitem{Hoehler84}
        G. H\"{o}hler, \emph{Pion Nucleon Scattering}, Part 2, Landolt-B\"ornstein:
        Elastic and Charge Exchange Scattering of Elementary Particles, Vol. 9b (Springer-Verlag, Berlin, 1983).
\bibitem{ConMap-Use} C. G. Boyd, B. Grinstein, and R. F. Lebed,  Phys. Rev. Lett. \textbf{74}, 4603 (1995); R. J. Hill, and G. Paz, Phys.
Rev. \textbf{D 82}, 113005 (2010).
\bibitem{Gribov} V.N. Gribov, 'Strong Interactions of hadrons at high energies', Cambridge U. Press, 2009.
\bibitem{Ceci2011} S. Ceci, M. D\"{o}ring, C. Hanhart, S. Krewald, U.-G. Mei\ss{}ner, and A. \v{S}varc, Phys. Rev.  \textbf{C 84},
015205 (2011).
\bibitem{MAID} D.~Drechsel, S.~S.~Kamalov and L.~Tiator, Eur.\ Phys.\ J.\ A {\bf 34}, 69 (2007);[http://www.kph.uni-mainz.de/MAID/].
\bibitem{Svarc2014}
 A. \v{S}varc, M. Had\v{z}imehmedovi\'{c}, R. Omerovi\'{c}, H. Osmanovi\'{c}, J. Stahov, arXiv:1401.1947 [nucl-th].
\bibitem{Hanstein:1996}
  O.~Hanstein, D.~Drechsel and L.~Tiator, Phys.\ Lett.\ B {\bf 385}, 45 (1996).
\bibitem{Workman:1999}
  R.~L.~Workman and R.~A.~Arndt,  Phys.\ Rev.\ C {\bf 59}, 1810 (1999).
\bibitem{Workman:2013rca}
  R.~L.~Workman, L.~Tiator and A.~Sarantsev,
  Phys.\ Rev.\ C {\bf 87}, 068201 (2013).
\bibitem{Kamano:2013}
  H.~Kamano, S.~X.~Nakamura, T.~-S.~H.~Lee and T.~Sato, Phys.\ Rev.\ C {\bf 88}, 035209 (2013).
\bibitem{Ronchen:2014}
  D.~R\"onchen, M.~D\"oring, F.~Huang, H.~Haberzettl, J.~Haidenbauer, C.~Hanhart, S.~Krewald and U.~-G.~Mei\ss{}ner {\it et al.},
  arXiv:1401.0634 [nucl-th].

\end{thebibliography}
\end{document}